\documentclass[twocolumn,showpacs,preprintnumbers,amsmath,amssymb,floatfix]{revtex4}

\usepackage{graphicx}
\usepackage{dcolumn}
\usepackage{bm}

\begin{document}


\title{Simultaneous measurement of the photodisintegration of 
$^{4}$He in the giant dipole resonance region}

\author{T. Shima}
 \email{shima@rcnp.osaka-u.ac.jp}
\author{S. Naito}
 \altaffiliation[Present address]{: Toshiba Co., Ltd., 8, 
Shinsugita-Cho, Isogo,Yokohama 235-8523, Japan}
\author{Y. Nagai}
\affiliation{Research Center for Nuclear Physics, 
Osaka University, Ibaraki, Osaka 567-0047, Japan}
\author{T. Baba}
 \altaffiliation[Present address]{: Hitachi, Ltd., 1-280, 
 Higashi-koigakubo, Kokubunji, Tokyo 185-8601, Japan}
\author{K. Tamura}
 \altaffiliation[Present address]{: ERNST\&YOUNG SHINNIHON, 
 Hibiya Kokusai Bldg., 2-2-3, Uchisaiwai-cho, 
 Chiyoda-ku, Tokyo 100-0011, Japan}
\author{T. Takahashi}
 \altaffiliation[Present address]{: Ricoh. Co., Ltd., 30-1, 
 Saho, Yashiro-cho, Kato-gun, Hyogo 673-1447, Japan}
\affiliation{Department of Physics, Tokyo Institute of 
Technology, Meguro, Tokyo 152-8551, Japan}
\author{T. Kii}
\author{H. Ohgaki}
\affiliation{Institute of Advanced Energy, Kyoto 
University, Uji, Kyoto 611-0011, Japan}
\author{H. Toyokawa}
\affiliation{Photonics Research Institute, National 
Institute of Advanced Industrial Science and Technology, 
Tsukuba, Ibaraki 305-8568, Japan}

\date{\today}

\begin{abstract}
We have performed for the first time the simultaneous measurement 
of the two-body and three-body photodisintegration cross-sections 
of $^{4}$He in the energy range from 21.8 to 29.8 MeV using 
monoenergetic pulsed photons and a 4$\pi$ time projection chamber 
containing $^{4}$He gas as an active target in an event-by-event mode. 
The photon beam was produced via the Compton backscattering of 
laser photons with high-energy electrons. The $^{4}$He($\gamma$,p)$^{3}$H 
and $^{4}$He($\gamma$,n)$^{3}$He cross sections were found to increase 
monotonically with energy up to 29.8 MeV, in contrast to the result of 
a recent theoretical calculation based on the Lorentz integral transform 
method which predicted a pronounced peak at around 26$-$27 MeV. The 
energy dependence of the obtained $^{4}$He($\gamma$,n)$^{3}$He cross 
section up to 26.5 MeV is marginally consistent with a Faddeev-type 
calculation predicting a flat pattern of the excitation function. The 
cross-section ratio of $^{4}$He($\gamma$,p)$^{3}$H to 
$^{4}$He($\gamma$,n)$^{3}$He is found to be consistent with the expected 
value for charge symmetry of the strong interaction within the 
experimental uncertainty in the measured energy range. The present 
results for the total and two-body cross-sections of the 
photodisintegration of $^{4}$He are compared to previous experimental 
data and recent theoretical calculations. 
\end{abstract}


\maketitle

\section{\label{sec:level1}Introduction}

The unique feature of $^{4}$He as the lightest self-conjugate nucleus 
with the simplest closed-shell structure prompts both experimentalists 
and theorists to study its photodisintegration reaction in the giant 
dipole resonance (GDR) region. Since the reaction proceeds mainly by 
an electric dipole (E1) transition in the GDR region, the 
photodisintegration study provides a wealth of fundamental information 
on nucleon-nucleon (NN) interactions, meson exchange currents 
\cite{Unkelbach} as well as the possibility to study the charge symmetry 
of the strong interaction \cite{Barker}. The photodisintegration study 
also gives important insight on the rapid neutron capture process 
(r-process) nucleosynthesis induced by neutrino-driven wind from a 
nascent neutron star \cite{Woosley}, since the neutrino transitions are 
the direct analogs of the giant electric dipole resonance observed in 
the photodisintegration \cite{Woosley,Meyer}. 

A considerable amount of theoretical work on the photodisintegration of 
$^{4}$He has been carried out in the GDR region. Above 50 MeV, the 
two-body $^{4}$He($\gamma$,p)$^{3}$H and $^{4}$He($\gamma$,n)$^{3}$He 
cross sections as well as the total cross section are well described by 
a plane-wave approximation, in which final state interactions (FSI) are 
known to play a minor role \cite{Sofianos}. Below 30 MeV, however, 
these cross sections are sensitive to FSI, meson exchange currents as 
well as to the choice of NN interaction \cite{Sofianos,Wachter}. 
Recently two different methods, one based on the Lorentz integral 
transform (LIT) \cite{Efros,Quaglioni} and another based on 
Faddeev-type Alt-Grassberger-Sandhas (AGS) integral equations 
\cite{Ellerkmann}, have been developed to accurately describe the 
low-energy dynamics of the $^{4}$He photodisintegration. Here it should 
be mentioned that although these models are quite different from each 
other, the calculated photodisintegration cross sections of $^{3}$H and 
$^{3}$He provided by these models agree with each other with high 
precision for the same NN interaction and three-nucleon forces (3NF) 
\cite{Golak}. However, the values of the photodisintegration cross 
section of $^{4}$He calculated by the same models differ 
significantly from each other. 
According to the calculation performed with the LIT method, both the 
total and two-body cross sections show a pronounced GDR peak at 
around 26$-$27 MeV, and the total cross section fully satisfies both 
the E1 sum rule and the inverse-energy-weighted E1 sum rule 
\cite{Efros,Quaglioni}. On the other hand, the calculation based on 
the AGS method, carried out for the $^{4}$He($\gamma$,n)$^{3}$He 
cross section, shows a flat pattern below the three-body threshold 
energy of 26.1 MeV, and the calculated cross section at 26.1 MeV is 
only about 60\% of the value derived by the LIT method 
\cite{Ellerkmann}. 

Experimentally the two-body, three-body, and total photodisintegration 
cross sections of $^{4}$He have been measured in the energy range from 
20 to 215 MeV using quasi-monoenergetic photon beams and/or 
bremsstrahlung photon beams. Concerning the two-body 
$^{4}$He($\gamma$,p)$^{3}$H and $^{4}$He($\gamma$,n)$^{3}$He reactions, 
their inverse, the nucleon capture reactions, were used to derive the 
photodisintegration cross sections. Previous data for the two-body and 
total cross sections are shown in Figs. 1(a), 1(b), and 1(c), respectively. 
It is quite interesting to note that above 35$-$40 MeV most of the 
previous $^{4}$He($\gamma$,p)$^{3}$H and $^{4}$He($\gamma$,n)$^{3}$He 
data agree with each other within their respective data sets 
\cite{Ellerkmann}. However, there appear to be discrepancies 
especially in the peak region of 25$-$26 MeV, where the data show 
either a pronounced GDR peak or a fairly flat excitation function 
as shown in Figs. 1(a) and 1(b). The experimental methods and their 
results in the previous measurements are briefly described below to 
obtain some hints of the origin of the large discrepancies mentioned 
above. Here, it would be quite interesting to 
note the discrepancies related to different photon probes. The 
$^{4}$He($\gamma$,p)$^{3}$H cross section, $\sigma(\gamma,p)$, was 
measured by detecting the protons by means of a NE213 liquid scintillator 
\cite{Bernabei88} and/or a Si(Li) detector array \cite{Hoorebeke}. 
Note that the latest result by Hoorebeke {\it et al.} using 34 MeV 
end-point bremsstrahlung photons \cite{Hoorebeke} is larger than the 
data by Bernabei {\it et al.} \cite{Bernabei88} using a monochromatic 
photon beam by about 40\% at around 30 MeV. The difference of these two 
data sets, however, becomes smaller with increasing the $\gamma$-ray 
energy, and they agree with each other at 33 MeV within the experimental 
uncertainty. The $^{3}$H(p,$\gamma$)$^{4}$He reaction cross section was 
measured using a tritium target absorbed into various metals by detecting 
a $\gamma$-ray by means of a NaI(Tl) detector [13-19]. Note that the latest 
result by Hahn {\it et al.} \cite{Hahn} is about 20\% larger than that by 
Feldman {\it et al.} \cite{Feldman}. In summary, the 
$^{4}$He($\gamma$,p)$^{3}$H cross section derived from both the 
photodisintegration and the inverse reaction shows a large discrepancy 
between different data sets, and the difference is quite large (about 50\%) 
at $E_{\gamma}=$ 25 MeV.

On the other hand, the $^{4}$He($\gamma$,n)$^{3}$He cross section, 
$\sigma(\gamma,n)$, was measured by detecting the neutrons with BF$_{3}$ 
neutron detectors and using bremsstrahlung photons \cite{Irish,Malcom} 
and/or monoenergetic photons \cite{Berman}. The results obtained using 
bremsstrahlung photons are larger by about 30$\sim$100\% than the result 
obtained using monoenergetic photons in the region between 25 and 30 MeV. 
Similarly to the case noted above, the difference between these data sets 
with different photon beams becomes smaller with increasing $\gamma$-ray 
energies, and they agree with each other at 35 MeV within an experimental 
uncertainty. The $^{3}$He(n,$\gamma$)$^{4}$He reaction cross section was 
measured by detecting $\gamma$-rays with a NaI(Tl) and/or a BGO detector 
\cite{Ward,Komar}, and their measured cross sections in the $\gamma$-ray 
energy range from 22 to 33 MeV agree with the $^{4}$He($\gamma$,n)$^{3}$He 
data by Berman {\it et al.} within the experimental uncertainty 
\cite{Berman}. 

Simultaneous measurements of the cross-sections for all reaction channels 
were performed by detecting charged fragments from the photodisintegration 
by means of cloud chambers using bremsstrahlung photon beams in the energy 
range from 21.5 to 215 MeV \cite{Gorbunov68,Gorbunov58}, from 20.5 to 
150 MeV \cite{Arkatov74,Arkatov70}, and from 24 to 46 MeV 
\cite{Balestra77,Balestra79}, respectively. The results obtained with these 
measurements are 30$\sim$70\% larger than the cross sections obtained with 
monoenergetic photon beams or tagged photon beams. 

The elastic photon scattering of $^{4}$He was performed in the energy range 
from 23 to 34 MeV to derive indirectly the total photodisintegration cross 
section of $^{4}$He \cite{Wells}. The results by Gorbunov {\it et al.} 
\cite{Gorbunov58} agree with those by Arkatov {\it et al.} \cite{Arkatov74} 
and also by Wells {\it et al.} \cite{Wells} within the experimental 
uncertainty (see Fig. 1(c)). 

\begin{figure*}
\includegraphics[width=.6\linewidth]{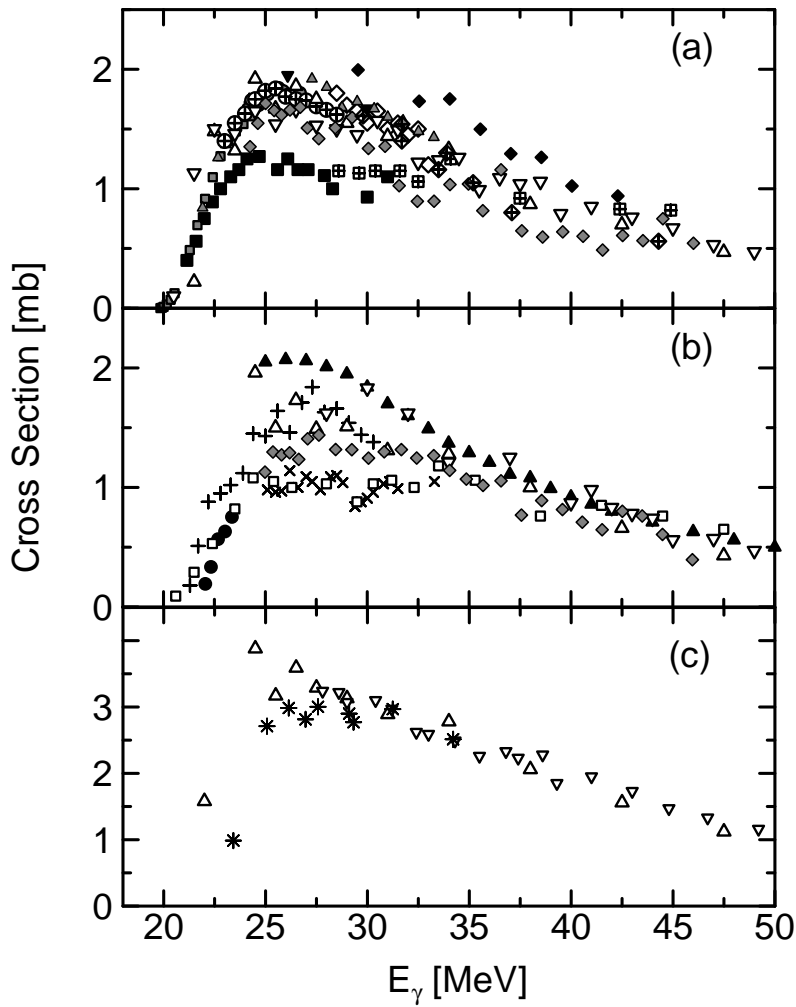}
\caption{\label{fig:Fig1}Available data of the 
$^{4}$He photodisintegration cross sections: 
(a) ($\gamma$,p) cross sections, 
gray circles; Gardner {\it et al.} \cite{Gardner}, 
crossed circles; Gemmell {\it et al.} \cite{Gemmell}, 
open upward triangles; Gorbunov \cite{Gorbunov68}, 
gray triangles; Meyerhof {\it et al.} \cite{Meyerhof}, 
open downward triangles; Arkatov {\it et al.} 
\cite{Arkatov74}, 
gray diamonds; Balestra {\it et al.} \cite{Balestra77}, 
filled diamonds; McBroom {\it et al.} \cite{McBroom}, 
filled downward triangles; Calarco {\it et al.} 
\cite{Calarco}, 
crossed squares; Bernabei {\it et al.} \cite{Bernabei88}, 
filled squares; Feldman {\it et al.} \cite{Feldman}, 
open diamonds; Hoorebeke {\it et al.} \cite{Hoorebeke}, 
gray squares; Hahn {\it et al.} \cite{Hahn}, 
(b) ($\gamma$,n) cross sections, 
open upward triangles; Gorbunov \cite{Gorbunov68}, 
crosses; Irish {\it et al.} \cite{Irish}, 
filled upward triangles; Malcom {\it et al.} \cite{Malcom}, 
open squares; Berman {\it et al.} \cite{Berman}, 
open downward triangles; Arkatov {\it et al.} \cite{Arkatov74}, 
gray diamonds; Balestra {\it et al.} \cite{Balestra77}, 
diagonal crosses; Ward {\it et al.} \cite{Ward}, 
filled circles; Komar {\it et al.} \cite{Komar}, 
(c) total photoabsorption cross sections, 
open upward triangles; Gorbunov {\it et al.} \cite{Gorbunov58}, 
open downward triangles; Arkatov {\it et al.} \cite{Arkatov74}, 
asterisks; Wells {\it et al.} \cite{Wells}. 
The error bars are not shown for clarity.}
\end{figure*}

The electromagnetic property of the photodisintegration 
cross-section of $^{4}$He in the giant resonance region has been 
discussed in terms of the electric dipole (E1) radiation 
\cite{Ellerkmann}. Experimentally below 26.6 MeV the E1 dominance with 
a small M1 contribution of less than 2\% has been shown by measuring 
angular distributions of cross-sections and/or analyzing powers for 
the inverse $^{3}$H(p,$\gamma$)$^{4}$He reaction 
\cite{McBroom,Calarco,Wagenaar}. Theoretically an E2 contribution to 
the total two-body cross-section is estimated to be small, about 6\%, 
even at $E_{\gamma} =$ 60 MeV \cite{Ellerkmann}.

The cross-section ratio of $^{4}$He($\gamma$,p)$^{3}$H to 
$^{4}$He($\gamma$,n)$^{3}$He, $R_{\gamma} = 
\sigma(\gamma,p)/\sigma(\gamma,n)$, in the GDR region has been 
used to test the validity of the charge symmetry of the strong 
interaction. When charge symmetry is valid, the ratio is 
about unity for pure E1 excitations \cite{Barker}. $R_{\gamma}$ 
has been obtained experimentally with values ranging from 1.1 to 
1.7 by separate measurements of $\sigma(\gamma,p)$ and 
$\sigma(\gamma,n)$ in the GDR region \cite{Calarco}. From 
simultaneous measurements of the $^{4}$He($\gamma$,p)$^{3}$H 
and $^{4}$He($\gamma$,n)$^{3}$He reactions using cloud 
chambers and bremsstrahlung photon beams, $R_{\gamma}$ was 
obtained as 1.0$\sim$1.5 in the energy range from 23 to 44 MeV 
\cite{Gorbunov68,Arkatov74,Balestra77}. Recently $R_{\gamma}$ of 
1.1 was obtained by a simultaneous ratio measurement of the 
$^{4}$He($\gamma$,p)$^{3}$H and $^{4}$He($\gamma$,n)$^{3}$He 
reactions in the energy range from 25 to 60 MeV 
\cite{Florizone}. The measurement was performed by detecting 
a charged fragment emitted at 90$^{\circ}$ with respect to 
an incident tagged photon beam direction by means of 
windowless $\Delta$E-E telescopes. Here an angular 
distribution effect of a fragment was corrected for using 
theory. 

In summary, although considerable experimental efforts have 
been made in determining $R_{\gamma}$, there remains a large 
discrepancy between separate measurements and simultaneous ratio 
measurements for the $^{4}$He($\gamma$,p)$^{3}$H and 
$^{4}$He($\gamma$,n)$^{3}$He channels. Hence, one can hardly 
discuss the validity of the charge symmetry of the strong nuclear 
force using existing data. Hence it is highly required to accurately 
measure these cross sections with use of a new method in the GDR 
region, in particular between 22 and 32 MeV 
\cite{Bernabei88,Hoorebeke,Wells,Vinokurov}. 

In designing a new experiment, it would be worthwhile to reconsider 
what we learned from previous data. Firstly, we notice that both the 
$^{4}$He($\gamma$,p)$^{3}$H and $^{4}$He($\gamma$,n)$^{3}$He cross 
sections measured with bremsstrahlung photons are much larger than 
those measured with monoenergetic photons in the energy range from 
22 to 30 MeV, and they agree with each other above $\sim$35 MeV. 
Theoretically the two-body as well as the total cross sections 
are well described by a plane-wave approximation and they agree 
with previous data above 50 MeV \cite{Sofianos}. Secondly, most 
experiments were performed separately for the 
$^{4}$He($\gamma$,p)$^{3}$H and $^{4}$He($\gamma$,n)$^{3}$He channels 
via the photodisintegration reactions and/or the inverse nucleon 
capture reactions. Thirdly, the simultaneous two-body and three-body 
cross section measurements were performed using a cloud chamber, which 
did not allow us to take data in an event-by-event mode with a pulsed 
photon beam, which is necessary to reject background. One may conclude 
that the large discrepancies between different data sets could be due 
to background inherent to incident photon beams and/or due to an 
uncertainty of the normalization of the $^{4}$He($\gamma$,p)$^{3}$H 
and $^{4}$He($\gamma$,n)$^{3}$He cross sections.

In the present study we have carried out the simultaneous 
measurement of the two-body and three-body $^{4}$He 
photodisintegration cross sections in the energy region 
between 21.8 and 29.8 MeV using a monoenergetic pulsed laser 
Compton backscattering photon beam by means of a newly 
developed 4$\pi$ time projection chamber containing $^{4}$He 
gas as an active target. 

\section{\label{sec:level2}Experiment}
\subsection{\label{sec:level2a}Experimental method}

The experiment was carried out using a pulsed Laser Compton 
backscattering (LCS) photon beam at the National Institute of 
Advanced Industrial Science and Technology (AIST). The charged 
fragments from the photodisintegration of $^{4}$He were detected 
by means of a time projection chamber (TPC). A schematic view 
of an experimental setup is shown in Fig. 2. 

\begin{figure}
\includegraphics[width=.9\linewidth]{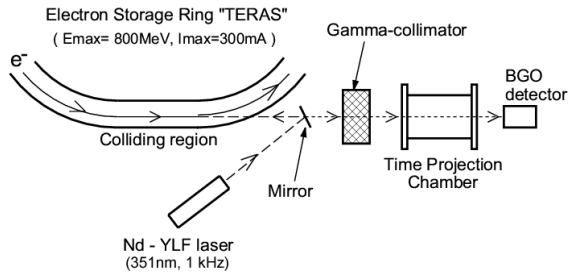}
\caption{\label{fig:Fig2} Experimental setup for measurement 
of the photodisintegration of $^{4}$He at AIST.}
\end{figure}

A quasi-monoenergetic pulsed LCS photon beam was produced 
via the Compton backscattering of the photons from a Nd:YLF 
laser in third harmonics ($\lambda=$ 351 nm) with electrons 
circulating in the 800 MeV storage ring TERAS at the AIST 
\cite{Ohgaki}. An LCS photon beam is well known to be an 
excellent probe to measure a photodisintegration cross 
section of a nucleus with little background associated with 
primary photon beam and with small uncertainty in determination 
of the LCS photon flux using a $\gamma$-ray detector. Even with 
this kind of setup, there are several difficulties inherent to the 
measurement of the photodisintegration cross section of $^{4}$He, 
among which the cross section is small (about $\sim$1 mb), the photon 
beam flux is low, the target density of $^{4}$He is low, and the 
energies of the fragments from the photodisintegration of $^{4}$He 
in the GDR region are quite low, typically less than a few MeV. 
Hence, it has been crucial to develop a new detector, which enabled 
us to make a simultaneous measurement of the two-body and three-body 
photodisintegration cross sections of $^{4}$He by detecting such a 
low energy fragment with an efficiency of 100\% with a large solid angle 
of 4$\pi$, and with a large signal-to-noise ratio. 

In the present study we constructed a TPC which meets the mentioned 
requirements. 

\subsection{\label{sec:level2b}Laser Compton backscattering 
(LCS) photon beam}

A pulsed LCS $\gamma$-ray with the maximum energies $E_{max}=$ 
22.3, 25, 28 and 32 MeV was used in the present experiment, 
obtained by changing the electron energy of the TERAS. The pulse 
width of the electron beam was 6 ns with a repetition rate of 100 MHz, 
while that of the laser photon beam was 150 ns with a repetition rate 
of 1 kHz. Pulsed laser photons scattered by electrons were collimated 
using a lead block with a hole of 2 mm in diameter and 200 mm in length 
to obtain quasi-monoenergetic LCS $\gamma$-rays. The absolute value of 
$E_{max}$ was determined with accuracy better than 1\% from the wavelength 
of the laser light and the kinetic energy of the electron beam. The 
electron beam energy has been calibrated by measuring the LCS $\gamma$-ray 
energy generated with Nd:YAG laser photons in fundamental mode 
($\lambda=$ 1064 nm) \cite{Ohgaki}. The half-width of the $\gamma$-ray 
energy distribution was 2.5 MeV at $E_{max}=$ 32 MeV, and the obtained 
$\gamma$-ray intensity was about 10$^{4}$ photons/s. The TPC was placed 
3 m downstream of the lead collimator. 

\subsection{\label{sec:level2c}Time projection chamber (TPC)}

A 4$\pi$ time projection chamber containing $^{4}$He gas as 
an active target was constructed to detect the charged fragments 
from the photodisintegration of $^{4}$He with an efficiency of 
100\%. The TPC was contained in a vessel with a size of 244 mm 
in inner diameter and 400 mm in length. A mixed gas of 80\% 
natural He and 20\% CH$_{4}$ with a total pressure of 1000 Torr 
was filled in the vessel as a target for the photon-induced 
reactions and an operational gas of the TPC.

The TPC consisted of a drift region with a uniform electric field 
with an area of 60$\times$60 mm$^{2}$ and a length of 250 mm, 
and a multi-wire proportional counter (MWPC) region as shown 
in Fig. 3. The MWPC consisted of one anode plane and two cathode 
planes, which were set with a gap of 2 mm. Each plane had 30 
wires with a spacing of 2 mm. In order to obtain two-dimensional 
track information of a charged fragment, cathode wires in front 
of and behind the anode plane were stretched along x- and y-axes, 
respectively. Here the x- and y-directions were defined to be 
parallel to and perpendicular to the anode wires, respectively. 

\begin{figure}
\includegraphics[width=.9\linewidth]{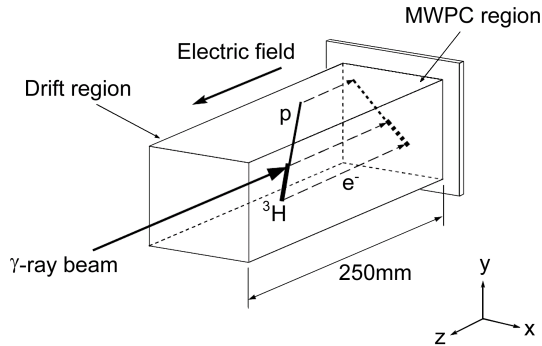}
\caption{\label{fig:Fig3} Schematic drawing of the 
structure of the TPC.}
\end{figure}

The TPC operates as follows. Electrons were produced by the 
interaction of a charged fragment with the mixed gas along the 
fragment path in the drift region. The electrons were drifted 
along the uniform electric field toward the MWPC region, where 
they were multiplied via an avalanche process. The avalanche 
signal was picked up with both the anode- and cathode-wires. 
The cathode signals were used to measure the track of a charged 
fragment on an x-y plane, since the directions of these cathode 
wires were perpendicular to each other. A z-position of a charged 
fragment was determined by measuring the drift time of the 
electrons with use of a time to digital converter as described 
below. An anode signal was used to determine the amount of energy 
loss of a fragment in the drift region of the TPC. Both, track 
and energy loss signals of a charged fragment were used to 
clearly identify a reaction channel. It should be noted 
that since a light charged fragment did not stop in the drift 
region, we observed various energy loss signals depending on 
a charged fragment type and on incident LCS $\gamma$-ray energy. 
An external magnetic field has not been used in the present TPC 
configuration. 

The performance of the TPC was studied using the $^{241}$Am 
$\alpha$-ray source and a Si detector. The energy resolution 
of the TPC was measured as being 7.5\% (FWHM) per anode wire. 
Since the energy measured by an anode wire depends on the 
emission angle of a fragment with respect to the anode wire 
direction, we collimated an $\alpha$-ray and determined its 
emission angle by using a coincidence signal between the TPC 
anode signal and the signal from the Si detector. A drift velocity 
of ionized electrons was measured as a function of the 
z-position using the same measuring system. A typical value of 
the drift velocity was 7.00$\pm$0.14 mm/$\mu$s. The time 
resolution was obtained as being 32 ns (1$\sigma$), which 
corresponded to the position resolution along the z-direction 
of 0.22 mm (1$\sigma$). Detailed description of the TPC will 
be published elsewhere \cite{Kii}. 

\subsection{\label{sec:level2d}Electronics and data acquisition}

A schematic diagram of data acquisition system is shown in 
Fig. 4. A linear signal from the preamplifier was used as a stop 
signal for a time to digital converter (TDC) after discriminating 
the electronic noise by a comparator. A common start signal for 
the TDC is obtained from the output of a pulsed laser clock. 
Both times, the leading edge and the trailing edge of an input signal 
are recorded on the TDC not only to determine the drift time of 
electrons but also to unambiguously identify the reaction 
channel. To measure the amount of energy loss of a charged fragment 
by integrating the current of a signal, we recorded its pulse 
shape using a flash ADC (FADC) and constructed a 
charge-integrated spectrum of a fragment in the off-line analysis. 

\begin{figure}
\includegraphics[width=.9\linewidth]{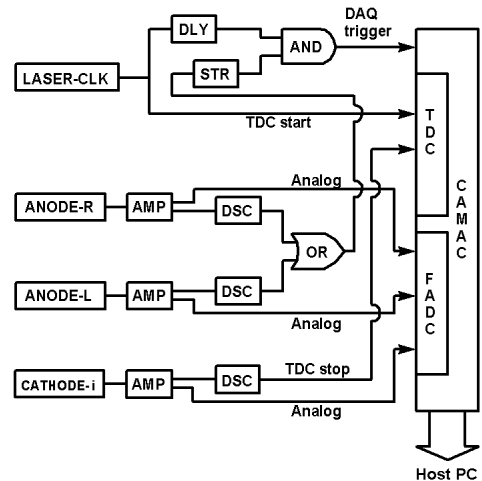}
\caption{\label{fig:Fig4} Block diagram of the data 
acquisition system. LASER-CLK; laser clock pulse, ANODE-R(L); 
sum of the linear signals from the anode wires in right- 
(left-) hand side with respect to the LCS photon beam axis, 
CATHODE-i; linear signal from the i-th cathode wire, AMP; 
preamplifier, DSC; discriminator, DLY; delay circuit, and 
STR; pulse stretcher.}
\end{figure}

Concerning the data acquisition, a trigger signal for the TPC is 
obtained from the clock pulse of the laser system. A logic signal 
from a cathode wire is sent to a discriminator to reject the noise 
signal and then sent to the TDC to measure a drift time of 
ionized electrons. An anode signal is used to generate a 
TPC-hit signal. When a 70 $\mu$s delayed signal of the laser clock 
pulse and the TPC-hit signal are in coincidence within a gate width 
of 100 $\mu$s, the data are acquired. The width is set longer than 
the maximum drift time (36 $\mu$s) of electrons in the TPC drift 
region in order to measure not only the photodisintegration event 
of $^{4}$He but also background events. Data from CAMAC modules are 
acquired by a personal computer and recorded on a hard disk drive in 
an event-by-event mode. The TPC count rate during the experiment was 
several tens of counts per second, and thus the dead time of the data 
taking system was a few \% (monitored during the measurement). 
A PC-based pulse-height analyzer was used for monitoring the LCS 
$\gamma$-ray intensities with a BGO detector as described below. 

\section{\label{sec:level3}ANALYSIS AND RESULTS}
\subsection{\label{sec:level3a}Event identification}

All pulse-height spectra taken by the FADC were analyzed 
to classify the observed events into photodisintegration events 
of $^{4}$He and $^{12}$C, and background events. We observed 
$^{12}$C events, since we used CH$_{4}$ gas. The total event 
rate in the present experiment was of several tens of counts 
per second, and the event rate from the photodisintegration 
of $^{4}$He and $^{12}$C was of several tens of counts per 
hour, less than one thousandth of the background events. A detailed 
description of electron background, natural background and 
photodisintegration events is given here below. 

\subsubsection{\label{sec:level3a1}Background events}

{\it(1) Electron events.} Most events taken by the FADC 
were due to background. The dominant background was 
originating from the interaction of LCS $\gamma$-rays with atomic 
electrons of $^{4}$He and CH$_{4}$ used for the TPC. Electron 
events were identified by their small pulse height. Note that 
electron energy loss rate in the TPC was small, of the order of 
0.1 keV/mm, since electron energy was in the range from a 
few MeV to several tens of MeV. Therefore most electron events 
could be discriminated by a discriminator. A typical spectrum 
taken by the FADC is shown in Fig. 5. Here a dotted line 
indicates a threshold level, which was set to further remove 
electron background during the off-line analysis. Typical 
tracks of electron events, which were detected with one 
anode trigger signal and whose energies were above the 
threshold level, are shown in Fig. 6. Note that we could see 
several tracks for one anode trigger signal. In addition, 
observed tracks were not straight, seldom crossed the LCS 
$\gamma$-ray axis, and their track width was quite thin. 
These features allowed us to unambiguously identify electron 
events.\\
\ \\
\begin{figure}
\includegraphics[width=.9\linewidth]{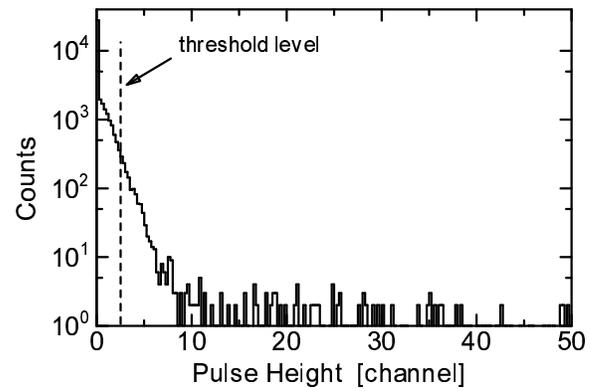}
\caption{\label{fig:Fig5} Pulse-height spectrum for all the 
acquired events. The huge component below $\sim$10 ch is 
mainly due to scattered electrons.}
\end{figure}

\begin{figure}
\includegraphics[width=.9\linewidth]{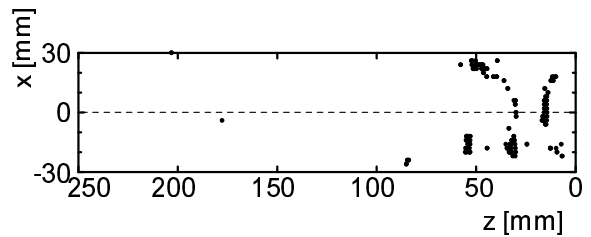}
\caption{\label{fig:Fig6} Example of a track of a 
scattered electron. The $\gamma$-ray beam is coming 
from the left-hand side. The dots indicate the envelopes 
of the electron clouds ionized by the scattered electrons. 
The dashed line denotes the incident $\gamma$-ray beam 
axis. The box is the drift region of the TPC (side view).}
\end{figure}
\ \\
{\it (2) Natural background events.} 
The natural background events are not correlated with 
the pulsed LCS $\gamma$-rays, and therefore the track did 
not cross the LCS $\gamma$-ray axis as shown in Fig. 7. 
Hence the natural background events could be clearly 
discriminated from the photodisintegration events of 
$^{4}$He and $^{12}$C. Since the track width of natural 
background is wider than that of electrons, the 
background might be due to an $\alpha$-particle from a 
natural radioactivity such as Rn contained in the TPC 
chamber or in the mixed gas of natural He and CH$_{4}$. 

\begin{figure}
\includegraphics[width=.9\linewidth]{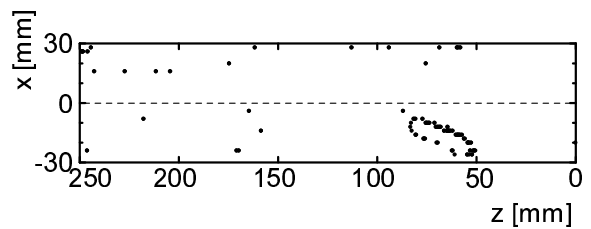}
\caption{\label{fig:Fig7} Example of a track of a natural 
background event.}
\end{figure}

\subsubsection{\label{sec:level3a2}Photodisintegration 
events of $^{4}$He and $^{12}$C}

Both electron and natural background events were identified 
as described above. Consequently, the background free (BF) 
events, which contained the photodisintegration events of 
$^{4}$He and $^{12}$C, were obtained from all the events 
recorded on the FADC. We checked the path length, the track 
width and the pulse height of each BF event to finally 
identify a reaction channel for the photodisintegration of 
$^{4}$He and $^{12}$C. 

The calculated path length of the various fragments from the 
photodisintegration of $^{4}$He and $^{12}$C in the present 
experiment are listed in Table I. The path length of a light 
fragment such as p, $^{3}$H and $^{3}$He is much longer than 
that of a heavy fragment such as $^{11}$B and $^{11}$C. Hence, 
the photodisintegration of $^{4}$He can be separated from that 
of $^{12}$C by referring to the path length of a charged fragment.

\begin{table}
\caption{\label{tab:table1}
Maximum ranges of the fragments from the photodisintegrations of 
$^{4}$He and $^{12}$C (unit:mm).}
\begin{ruledtabular}
\begin{tabular}{ccccccc}
Reaction & $Q$ & Fragment & 
\multicolumn{4}{c}{$E_{\gamma}$
[MeV]}\\ \cline{4-7}
channel & [MeV] & & 22.3 & 25 & 28 & 32 \\ \hline
$^{4}$He($\gamma$,p)$^{3}$H & -19.81 & p
& 130 & 580 & 1310 & 2690 \\
 & & $^{3}$H & 14.5 & 45 & 91 & 174 \\ \hline
$^{4}$He($\gamma$,n)$^{3}$He & -20.58 & $^{3}$He
& 6 & 17 & 58 & 135 \\ \hline
$^{4}$He($\gamma$,pn)$^{2}$H & -26.07 & p
& $-$ & $-$ & 84 & 590 \\
 & & $^{2}$H & $-$ & $-$ & 34 & 214 \\ \hline
$^{12}$C($\gamma$,p)$^{11}$B & -15.96 & p
& 1080 & 2230 & 3770 & 6440 \\
 & & $^{11}$B & 4.3 & 6 & 7 & 9 \\ \hline
$^{12}$C($\gamma$,n)$^{11}$C & -18.72 & $^{11}$C
& 3 & 4.3 & 5.5 & 7.4
\end{tabular}
\end{ruledtabular}
\end{table}

The track width of a charged fragment was obtained by 
converting both times of the leading edge and the trailing 
edge of a cathode signal into the z-coordinate of the 
fragment track. Note that as the pulse height of a cathode 
signal becomes higher, the time difference between these two 
edges is larger, and thus the track width becomes wider. 
Hence, the track width of a charged fragment provides 
energy loss information of a fragment in the TPC. Since the 
energy loss rate of a fragment depends on the fragment type 
(p, $^{2}$H, $^{3}$H, $^{3}$He, $^{4}$He, $^{11}$B, and 
$^{11}$C), the track width of a fragment was used to identify 
its photodisintegration reaction channel together with the 
path length, the charge-integrated pulse height taken by 
the FADC and the reaction kinematics. 

The measured pulse height spectrum of a fragment was compared 
to the spectrum calculated by a Monte-Carlo method. The 
Monte-Carlo calculation simulated the kinematics of the 
photodisintegration events, the migration of drift electrons, 
and the pulse shapes of the signals from the anode and cathode 
wires. The calculation has been performed as follows. 
Firstly, an incident intrinsic LCS photon spectrum of given energy 
was generated which reproduced a measured energy spectrum with 
a NaI(Tl) detector. Then, a reaction point was randomly chosen 
in the region irradiated by the LCS photon beam in the TPC. 
The track of a charged fragment emitted by a photodisintegration 
reaction of $^{4}$He and/or $^{12}$C was calculated by considering 
the LCS photon energy and the Q-value of the reaction. In order 
to calculate the emission angle of a charged fragment we assumed 
an E1 angular distribution and an isotropic distribution for the 
two-body and three-body channels of the photodisintegration of 
$^{4}$He, respectively. Note that the E1 dominance of the two-body 
channel was experimentally shown as mentioned above 
\cite{McBroom,Calarco,Wagenaar}, and the isotropic fragment 
distribution was also observed for the $\gamma$-ray energy range 
from 28 to 60 MeV within the experimental uncertainty 
\cite{Gorbunov58,Arkatov70,Balestra79}. The energy deposited by 
a charged fragment was calculated as a function of the distance 
from a reaction point using the energy loss formula given by 
Ziegler {\it et al.} \cite{Ziegler}, and was converted to the 
number of ionized electrons using the ionization energy of electrons 
in the TPC gas. The drift time of ionized electrons was calculated 
using the local drift velocity, which has been obtained as a 
function of the electric field strength in the TPC as described 
above. Using the drift time thus calculated, the shaping time of 
an amplifier, and the threshold level of a discriminator, we 
obtained the simulated data of FADC and TDC for each wire. 
The event data thus obtained were recorded and analyzed with 
the same procedure as for the data of the real measurements.\\
\ \\
{\it (1) Two-body channel of $^{4}$He photodisintegration.} 
This channel is characterized by the fact that two fragments 
p (n) and $^{3}$H ($^{3}$He) are emitted in the opposite 
direction with respect to the LCS $\gamma$-ray beam direction, 
with equal momentum in the center-of-mass system. This channel 
can be separated from the two-body channel of $^{12}$C disintegration 
by the completely different path lengths of the charged fragments as 
mentioned above.\vspace{3mm}\\
(i) $^{4}$He($\gamma$,p)$^{3}$H channel\vspace{1mm}\\
\hspace*{5mm}Both the proton and the triton were detected by the 
TPC. An event, which meets the reaction kinematics conditions 
mentioned above, is selected as a candidate event of the 
$^{4}$He($\gamma$,p)$^{3}$H event. Since the energy loss of 
$^{3}$H is a few times larger than that of a proton, the track 
width of $^{3}$H is wider than that of a proton. A typical track of 
an event observed at $E_{max} =$ 28 MeV consistent with the 
above-mentioned feature is shown in Fig. 8(a). 
The sum spectrum of the measured pulse height of p and of 
$^{3}$H is in good agreement with that of a Monte-Carlo simulation 
as shown in Fig. 8(b). This event can be unambiguously assigned as 
a $^{4}$He($\gamma$,p)$^{3}$H event.\vspace{3mm}\\
(ii) $^{4}$He($\gamma$,n)$^{3}$He channel\vspace{1mm}\\
\hspace*{5mm}The TPC was not sensitive to neutrons, and 
therefore only the $^{3}$He, which crossed the LCS $\gamma$-ray 
axis, was detected for this reaction channel. A typical track of 
the event observed at $E_{max} =$ 28 MeV is shown in Fig. 9(a). 
The track of $^{3}$He is shown to extend to the opposite side across 
the central axis of the TPC. This is due to the finite size of the 
LCS photon beam and the diffusion of secondary electrons during 
the migration to the MWPC. The pulse height spectrum of $^{3}$He 
agrees nicely with a simulated one as shown in Fig. 9(b). Note that 
the track length of $^{3}$He is much longer compared to that of 
$^{11}$C as shown in Fig. 10 (also Fig. 12(a)), and therefore the 
$^{4}$He($\gamma$,n)$^{3}$He events can be clearly separated 
from the $^{12}$C($\gamma$,n)$^{11}$C events.\vspace{3mm}\\
(iii) $^{4}$He($\gamma$,d)$^{2}$H channel\vspace{1mm}\\
\hspace*{5mm}We did not observe any candidate of the 
$^{4}$He($\gamma$,d)$^{2}$H reaction. Note that the 
$^{4}$He($\gamma$,d)$^{2}$H cross section was measured to be 
about 3.2 $\mu$b at the peak of $E_{\gamma} =$ 29 MeV \cite{Weller}, 
and it is therefore much smaller compared to the 
$^{4}$He($\gamma$,p)$^{3}$H and $^{4}$He($\gamma$,n)$^{3}$He 
cross sections (a few mb at the corresponding $\gamma$-ray 
energy). \\

\begin{figure}[h]
\includegraphics[width=.9\linewidth]{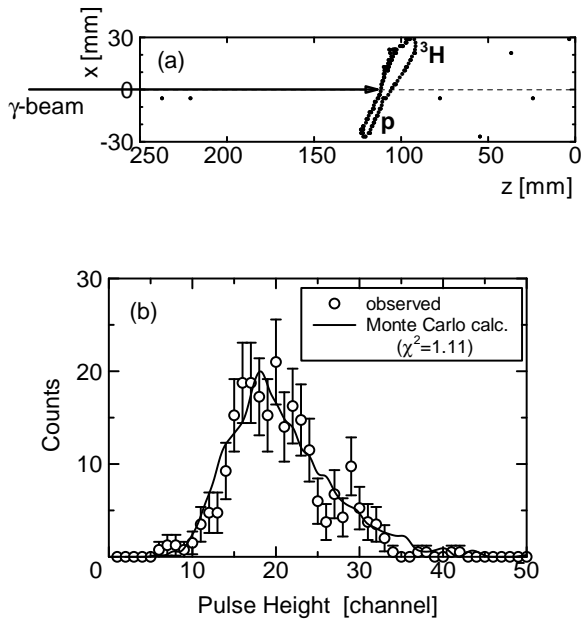}
\caption{\label{fig:Fig8} (a) Example of the 
$^{4}$He($\gamma$,p)$^{3}$H event.
(b) Total pulse height spectrum of the 
$^{4}$He($\gamma$,p)$^{3}$H reaction: open circles; 
observed, solid curve; fitting spectrum calculated with 
a Monte-Carlo simulation.}
\end{figure}

\ \vspace{3mm}\\

\begin{figure}
\includegraphics[width=.9\linewidth]{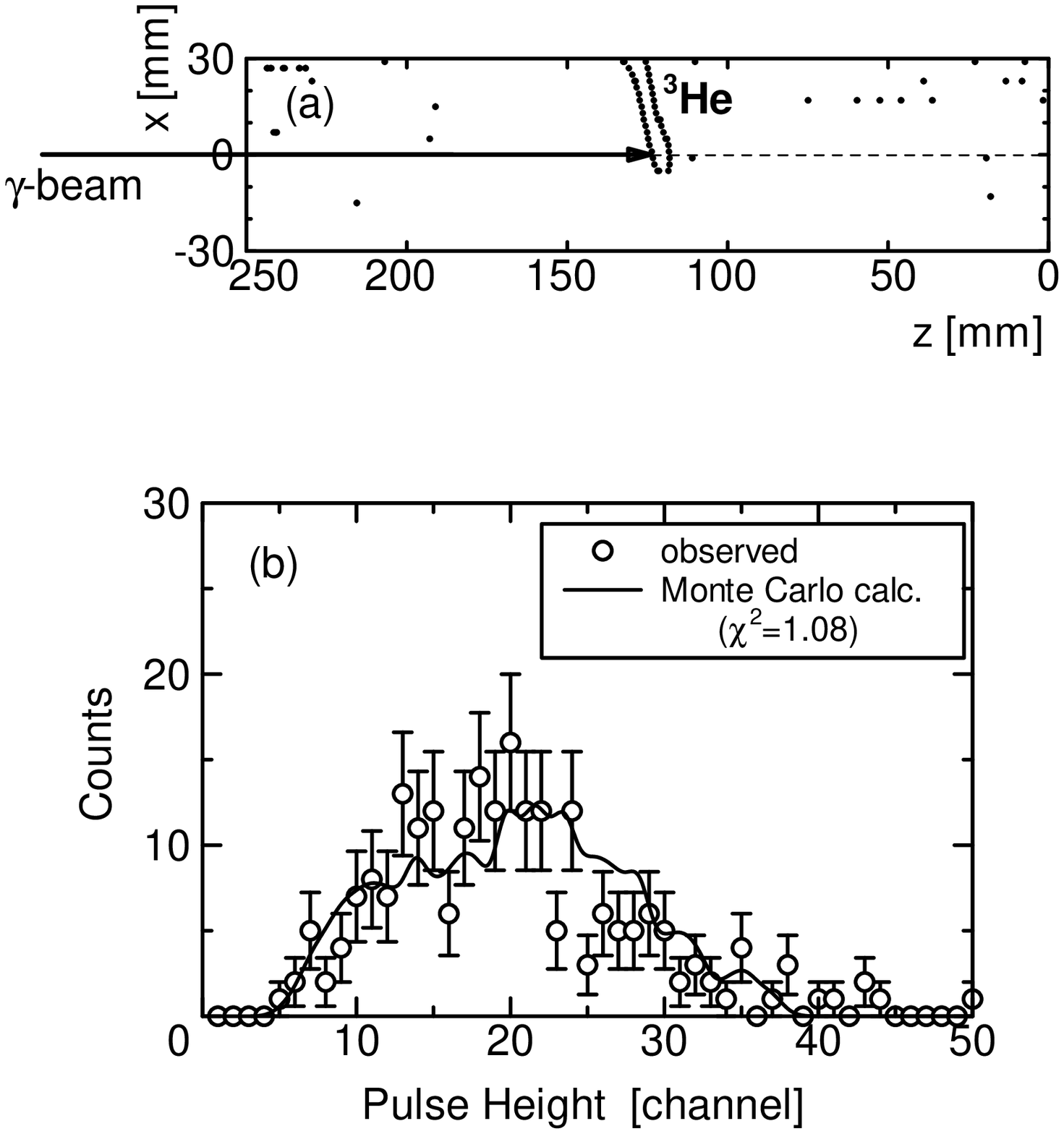}
\caption{\label{fig:Fig9} (a) Example of the 
$^{4}$He($\gamma$,n)$^{3}$He event. 
(b) Total pulse height spectrum of the 
$^{4}$He($\gamma$,n)$^{3}$He reaction.}
\end{figure}

\begin{figure}
\includegraphics[width=.9\linewidth]{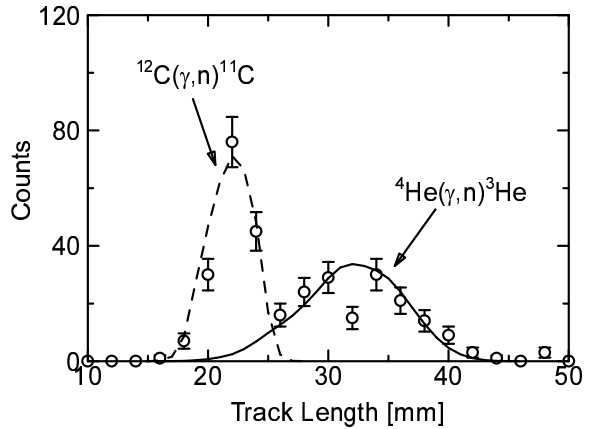}
\caption{\label{fig:Fig10} Distributions of the track length 
of charged fragments from the $^{4}$He($\gamma$,n)$^{3}$He 
and $^{12}$C($\gamma$,n)$^{11}$C reactions observed for 
$E_{max} =$ 28 MeV. The open circles are the experimental data. 
The solid curve and the dashed curve are the results of 
Monte-Carlo simulations for $^{4}$He($\gamma$,n)$^{3}$He 
and $^{12}$C($\gamma$,n)$^{11}$C, respectively.}
\end{figure}

\ \vspace{3mm}\\
%
%
\ \\
{\it (2) Two-body channel in the $^{12}$C photodisintegration.} 
This channel is characterized by the fact that two fragments 
p (n) and $^{11}$B ($^{11}$C) are emitted in the opposite 
direction with respect to the LCS $\gamma$-ray beam axis 
with equal momentum in the center-of-mass system. The path 
length of $^{11}$B and $^{11}$C, however, are much shorter 
than that of $^{3}$H and $^{3}$He, and therefore this 
two-body channel can be clearly separated from that of 
$^{4}$He.\vspace{3mm}\\
(i) $^{12}$C($\gamma$,p)$^{11}$B channel\vspace{1mm}\\
\hspace*{5mm}Both the proton and $^{11}$B are detected by 
the TPC. A typical track of an event which meets the 
above-mentioned condition, observed at $E_{max} =$ 28 MeV, 
is shown in Fig. 11(a). The path length of the proton is much 
longer than that of $^{11}$B, and the track width of the proton 
is much narrower than that of $^{11}$B. The sum spectrum of the 
measured pulse height of $^{11}$B and p is in good agreement 
with the Monte-Carlo simulation as shown in 
Fig. 11(b).\vspace{3mm}\\
(ii) $^{12}$C($\gamma$,n)$^{11}$C channel\vspace{1mm}\\
\hspace*{5mm}Only the track of $^{11}$C, which crossed the 
LCS $\gamma$-ray beam axis, was observed for this reaction channel. 
A typical track of a $^{12}$C($\gamma$,n)$^{11}$C event observed 
at $E_{max} =$ 28 MeV is shown in Fig. 12(a). The path length of 
$^{11}$C is much shorter than that of $^{3}$He as shown in Fig. 10, 
and therefore we could unambiguously discriminate the 
$^{12}$C($\gamma$,n)$^{11}$C events from those of the 
$^{4}$He($\gamma$,n)$^{3}$He reaction. The pulse height spectrum 
of $^{11}$C also agrees nicely with the simulated one as shown 
in Fig. 12(b).\\

\begin{figure}[t]
\includegraphics[width=.9\linewidth]{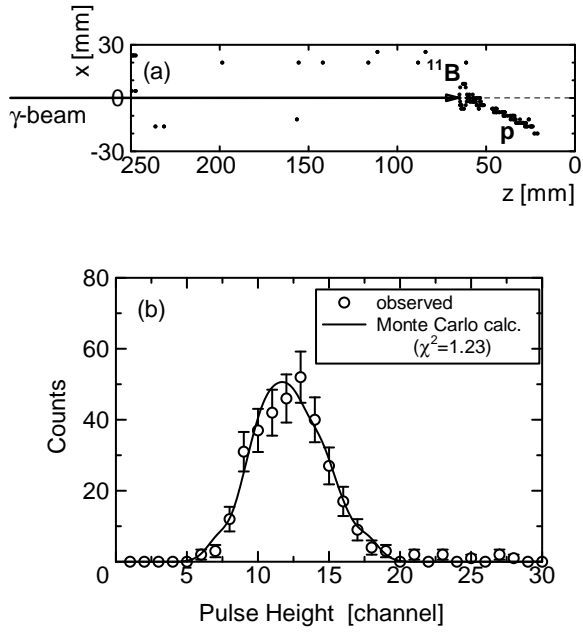}
\caption{\label{fig:Fig11} (a) Example of the 
$^{12}$C($\gamma$,p)$^{11}$B event. 
(b) Total pulse height spectrum of 
the $^{12}$C($\gamma$,p)$^{11}$B reaction.}
\end{figure}

\ \vspace{1mm}\\

\begin{figure}[h]
\includegraphics[width=.9\linewidth]{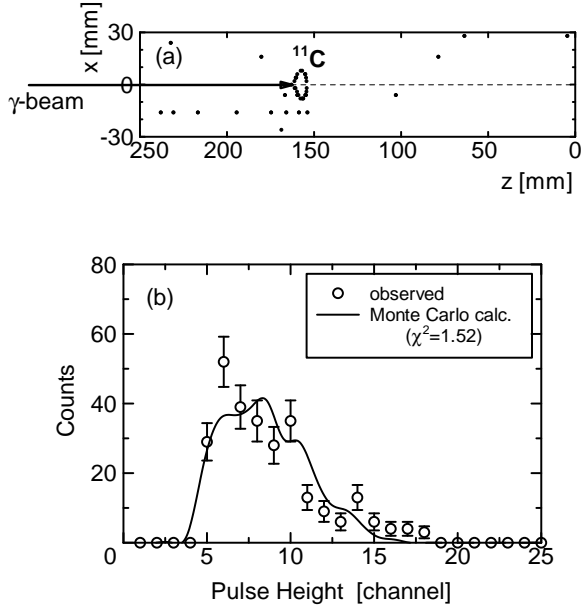}
\caption{\label{fig:Fig12} (a) Example of the 
$^{12}$C($\gamma$,n)$^{11}$C event. 
(b) Total pulse height spectrum of 
the $^{12}$C($\gamma$,n)$^{11}$C reaction.}
\end{figure}

\ \vspace{3mm}\\
{\it (3) Three-body channels.} \vspace{2mm}\\
(i) $^{4}$He($\gamma$,pn)$^{2}$H channel\vspace{1mm}\\
\hspace*{5mm}The Q-value of the $^{4}$He($\gamma$,pn)$^{2}$H 
reaction is -26.1 MeV, and therefore the reaction events could 
only be observed at $E_{max} =$ 28 and 32 MeV. This reaction event 
can be discriminated from that of the $^{4}$He($\gamma$,p)$^{3}$H 
reaction, because the tracks of the proton and deuteron are 
randomly oriented with respect to one another in the 
center-of-mass system and with respect to the LCS $\gamma$-ray 
beam axis, and the path length of the proton from the 
$^{4}$He($\gamma$,pn)$^{2}$H reaction is much shorter than that 
of the $^{4}$He($\gamma$,p)$^{3}$H reaction. A typical track of 
p and $^{2}$H observed at $E_{max} =$ 28 MeV is 
shown in Fig. 13. In this case, the track width of the proton 
is not constant and depends on the proton energy.

\begin{figure}[h]
\includegraphics[width=.9\linewidth]{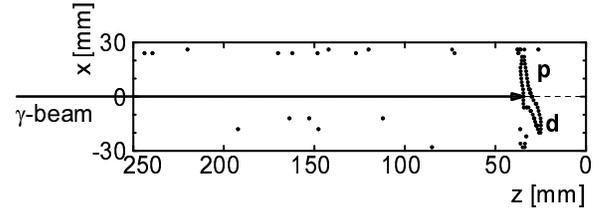}
\caption{\label{fig:Fig13} Example of the 
$^{4}$He($\gamma$,pn)$^{2}$H three-body event.}
\end{figure}

\ \vspace{3mm}\\
(ii) $^{12}$C($\gamma$,2$\alpha$)$^{4}$He channel
\vspace{1mm}\\
\hspace*{5mm}This reaction event can be easily identified by 
three tracks of the particles as shown in Fig. 14, which was 
observed at $E_{max} =$ 28 MeV.

\begin{figure}[h]
\includegraphics[width=.9\linewidth]{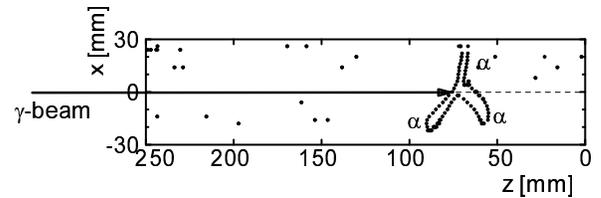}
\caption{\label{fig:Fig14} Example of the 
$^{12}$C($\gamma$,2$\alpha$)$^{4}$He event.}
\end{figure}

\ \\
\ \\
{\it (4) Photodisintegration reaction of $^{4}$He and/or $^{12}$C 
without LCS photon beams.} 
We checked for possible photodisintegration events of $^{4}$He 
and/or $^{12}$C caused by bremsstrahlung photons from the 
TERAS, but not LCS photons. Since such an event would 
occur continuously, the data corresponding to the drift time 
of between 40 and 64 $\mu$s were analyzed. We did not find 
any event which could be identified as any of the reaction channels 
of the photodisintegrations of $^{4}$He and/or $^{12}$C.\\

\subsection{\label{sec:level3b}Cross sections of the 
photodisintegrations of $^{4}$He ($\sigma_{i}(E_{\gamma})$)}

The incident LCS $\gamma$-ray has a finite energy spread, 
and the TPC efficiency depends on the incident $\gamma$-ray 
energy as described below. Hence, a partial cross section 
$\sigma_{i}(E_{\gamma})$ corresponding to the two-body 
and/or the three-body photodisintegration of $^{4}$He at 
a $\gamma$-ray energy $E_{\gamma}$ is given as follows:
\begin{equation}
Y_{i}=N_{t} \cdot L \cdot \Phi \times 
\frac{\int_{0}^{E_{max}} \varepsilon_{i}(E_{\gamma}) \cdot 
\sigma_{i}(E_{\gamma}) \cdot \phi(E_{\gamma}) dE_{\gamma}}
{\int_{0}^{E_{max}} \phi(E_{\gamma}) dE_{\gamma}}
\label{eq:one}.
\end{equation}

Here $Y_{i}$, $N_{t}$ and $L$ stand for the yield of 
a reaction channel $i$, the number density of the target 
nuclei, and the effective length of the TPC, respectively. 
$E_{max}$ denotes the maximum energy of the incident LCS 
$\gamma$-ray. 
The parameter $\varepsilon_{i}(E_{\gamma})$ is the detection 
efficiency of the TPC for a fragment emitted by the 
photodisintegration process at the $\gamma$-ray energy $E_{\gamma}$. 
The parameter $\phi(E_{\gamma})$ denotes the intensity of 
the incident LCS $\gamma$-ray at the energy $E_{\gamma}$. 
$\Phi$ is the incident LCS $\gamma$-ray flux, and is equal 
to the energy-integrated value of $\phi(E_{\gamma})$. 
The average cross section $<\sigma_{i}>$ and the 
weighted-mean reaction energy $<E_{\gamma}>_{i}$ are 
defined as

\begin{eqnarray}
<\sigma_{i}>=\frac{\int_{0}^{E_{max}} \varepsilon_{i}(E_{\gamma}) 
\cdot \sigma_{i}(E_{\gamma}) \cdot \phi(E_{\gamma}) dE_{\gamma}}
{\int_{0}^{E_{max}} \varepsilon_{i}(E_{\gamma}) \cdot 
\phi(E_{\gamma}) dE_{\gamma}}\nonumber \\
= \frac{Y_{i}}
{N_{t} \cdot L \cdot \int_{0}^{E_{max}} \varepsilon_{i}(E_{\gamma}) 
\cdot \phi(E_{\gamma}) dE_{\gamma}}, 
\label{eq:two}
\end{eqnarray}
\begin{equation}
<E_{\gamma}>_{i}=\frac{\int_{0}^{E_{max}} E_{\gamma} \cdot 
\varepsilon_{i}(E_{\gamma}) \cdot \sigma_{i}(E_{\gamma}) 
\cdot \phi(E_{\gamma}) dE_{\gamma}}
{\int_{0}^{E_{max}} \varepsilon_{i}(E_{\gamma}) \cdot 
\sigma_{i}(E_{\gamma}) \cdot \phi(E_{\gamma}) dE_{\gamma}}
\label{eq:three}.
\end{equation}

The parameters $\varepsilon_{i}$, $N_{t}$, $L$, $\Phi$ and 
$\phi$ were determined as discussed in the following subsections. 

\subsubsection{\label{sec:level3b1}Effective length ($L$) 
and detection efficiency ($\varepsilon_{i}(E_{\gamma})$) 
of the TPC}

Any charged fragments produced by the photodisintegration 
of $^{4}$He and/or $^{12}$C in the TPC produces electrons 
by interacting with atomic electrons in the He and CH$_{4}$ 
mixed gas in the TPC. Since the signal of an electron is picked up 
by the anode and the cathode wires, the efficiency 
$\varepsilon_{i}(E_{\gamma})$ of the TPC is expected to be as 
high as 100\% along the TPC geometrical drift length of 250 mm. 
However, since the electric field strength applied in the drift 
region is not uniform at both edges of the drift region, the 
efficiency is not constant in the whole length of the drift 
region. Hence, we measured a pulse height spectrum of 
$\alpha$-particles from the decay of $^{241}$Am to 
determine an effective length $L$, in which a pulse height 
was constant to provide a constant efficiency. The length 
$L$ is defined as 225 mm in the region between $z =$ 12.5 mm 
and 237.5 mm. Fig. 15 shows the position distribution of the 
photodisintegration event of $^{4}$He along the z-direction. 
It is clearly seen that the distribution is uniform within the 
effective length within the experimental uncertainty. 

\begin{figure}
\includegraphics[width=.9\linewidth]{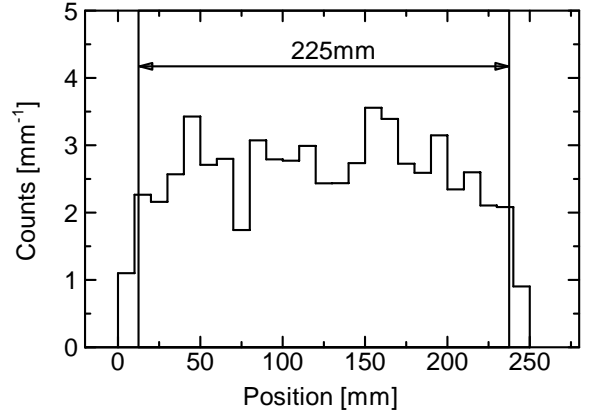}
\caption{\label{fig:Fig15} Position distribution of 
the $^{4}$He photodisintegration yield along the 
z-direction.}
\end{figure}

One might expect a 100\% efficiency $\varepsilon_{i}(E_{\gamma})$ 
within the effective length. However since the energy of a fragment 
from the photodisintegration of $^{4}$He and/or $^{12}$C is low 
and discriminators were used to reject electric noise of 
both the anode and cathode signals and electron background, 
it is necessary to investigate a possible change of the 
efficiency due to threshold levels of the discriminators of the 
cathode and anode signals. 

It should be mentioned that the anode signal was obtained 
by summing signals from several anode wires on either 
right-hand side or left-hand side with respect to the LCS 
$\gamma$-ray beam axis as shown in the block diagram of the 
data acquisition system in Fig. 4. Since the average energy 
deposit by a charged fragment from the photodisintegration 
is above 500 keV, an average pulse height of the summed 
anode signal is above 500 keV. A threshold level of the 
anode signal was set at about 5 keV by referring to the 
$\alpha$-ray pulse height spectrum of $^{241}$Am not to 
decrease the efficiency. 

On the other hand, a cathode signal was obtained from each 
cathode wire as shown in Fig. 4. A threshold level of 
the cathode signal was set at about 0.8 keV, and the effect 
of the discriminator on the efficiency was studied by 
making a pulse peak height spectrum of its signal. 
The spectrum was obtained by taking the maximum peak of all 
signals of a cathode wire taken by FADC. A typical spectrum 
for a proton and a $^{3}$H from the $^{4}$He($\gamma$,p)$^{3}$H 
channel is respectively shown in Fig. 16, where a solid curve 
indicates a peak height spectrum calculated by a Monte-Carlo 
method, and the dotted line is the threshold level set in the 
present measurement. The measured peak height spectrum is in 
good agreement with the simulated one both for the proton as 
well as for the $^{3}$H. Since the pulse peak height of the 
proton is well above the threshold level, the discriminator 
for the cathode signal does not decrease the TPC efficiency. 
Using the Monte-Carlo simulation, the efficiency 
$\varepsilon_{i}(E_{\gamma})$ is obtained as being 
0.97(5)$\sim$1.00(1) depending on the reaction energy. 
Here the bracket indicates the uncertainty of 
$\varepsilon_{i}(E_{\gamma})$, which was obtained by fitting 
a measured pulse-height spectrum with the simulated one. 

\begin{figure}
\includegraphics[width=.9\linewidth]{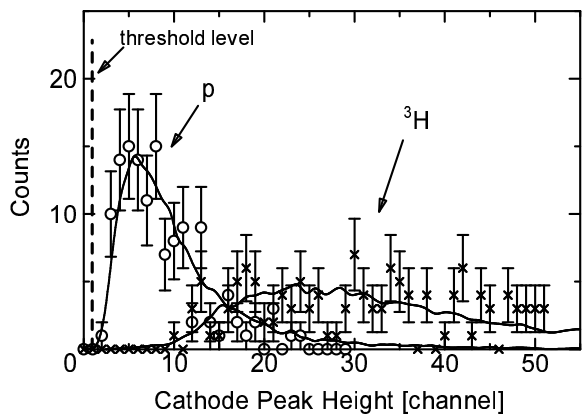}
\caption{\label{fig:Fig16} Peak pulse-height spectra 
of proton (open circles) and $^{3}$H (diagonal crosses) 
from the $^{4}$He($\gamma$,p)$^{3}$H reaction at 
$E_{max} =$ 28 MeV. Solid curves are the spectra calculated 
by a Monte-Carlo method.}
\end{figure}

\subsubsection{\label{sec:level3b2}Target number density 
($N_{t}$)}

The target number density $N_{t}$ was determined from 
measured pressure $P$, temperature $T$ and chemical purity 
(99.999\%) of the $^{4}$He gas in the TPC. The uncertainty 
in the determination of $N_{t}$ was evaluated to be 0.18\% 
due to the uncertainty in the determination of $P$ and $T$.

\subsubsection{\label{sec:level3b3}Incident LCS $\gamma$-ray 
flux ($\Phi$)}

The incident LCS $\gamma$-rays were measured using a BGO detector 
with a diameter of 50.8 mm and a length of 152.4 mm. A typical 
measured $\gamma$-ray spectrum is shown in Fig. 17, in which we 
see multiple peaks due to pile-up effects. 

\begin{figure}
\includegraphics[width=.9\linewidth]{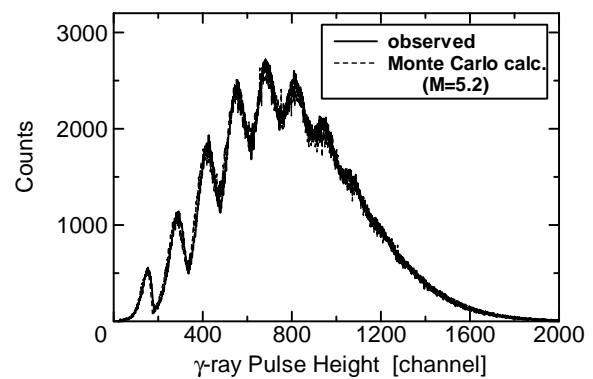}
\caption{\label{fig:Fig17} Typical $\gamma$-ray pulse 
height spectrum for $E_{max} =$ 28 MeV. The solid curve 
and the dashed curve represent the measured one and the 
Monte-Carlo simulation assuming an average photon 
multiplicity $M=$ 5.2, respectively.}
\end{figure}

The laser photon beam was pulsed with a pulse width of 150 ns 
and a repetition rate of 1 kHz, while the electron beam was 
also pulsed with a width of 6 ns and a repetition rate of 
100 MHz. Laser photons, therefore, collide several times with 
electron bunches circulating in the TERAS within one long 
laser pulse width, and LCS $\gamma$-rays with the same energy 
distribution were produced within a time interval of 150 ns. 
Note that this time interval was too short for the BGO system to 
decompose the multiple LCS $\gamma$-rays into an individual 
LCS $\gamma$-ray produced by one electron pulse. Consequently, 
the multiple $\gamma$-ray peaks were produced as pile-ups in the 
LCS $\gamma$-ray spectrum (see Fig. 17). 

The photodisintegration yield of $^{4}$He is proportional 
to an averaged number $M$ of multiple LCS $\gamma$-rays 
per laser pulse. The number $M$ was obtained by comparing 
a measured BGO spectrum to a calculated one obtained from a 
Monte-Carlo simulation \cite{Toyokawa}. The calculated spectrum 
was obtained with the following assumptions. The LCS $\gamma$-ray 
yield is proportional to the number of electrons (electron currents) 
times the number of laser photons. The probability density for 
generating LCS $\gamma$-rays per laser pulse is so small that the 
LCS $\gamma$-ray yield can be assumed to follow a Poisson 
distribution. The electron beam in the TERAS 
can be assumed to be a continuous beam because its repetition rate 
is much higher than that of the laser photon beam. The observed 
multiple peaks of the LCS $\gamma$-ray spectrum are assumed 
to be the sum of the pulse height spectra of each LCS $\gamma$-ray. 
This assumption is reasonable since the BGO responds to each 
$\gamma$-ray independently. The response function of the BGO 
detector to the LCS $\gamma$-ray was obtained by measuring 
the $\gamma$-ray spectrum with low flux, which was free from 
multiple peaks. Finally, the pulse shape of the BGO detector 
for multiple LCS $\gamma$-rays was obtained using both the 
time distribution of the LCS $\gamma$-ray measured by a 
plastic scintillation counter and a shaping time of 1 $\mu$s 
of an amplifier used for the BGO detector system. Based on 
these assumptions, a response function of the BGO detector with 
an averaged number $M$ of multiple LCS $\gamma$-rays was calculated 
by a Monte-Carlo method, and the number $M$ was obtained by fitting 
a measured spectrum with the multiple peaks with the calculated 
response function. A typical measured spectrum is in good agreement 
with the calculated one with $M =$ 5.2 as shown in Fig. 17. 

Using the number $M$ thus determined the LCS $\gamma$-ray 
total flux $\Phi$ is obtained as follows:

\begin{equation}
\Phi = M \times f \times T_{L}
\label{eq:four}.
\end{equation}

Here $f$ is a frequency of the laser pulse, and $T_{L}$ 
is a live time of the measurement. 
A $\gamma$-ray flux thus obtained has an uncertainty of about 2\%, 
which consists of statistics of the LCS $\gamma$-ray 
yield, an uncertainty of the response function of the 
BGO detector, and an uncertainty in the least-square 
fitting of the LCS $\gamma$-ray spectrum with multiple 
peaks measured with the BGO detector. 

It is interesting to see a relation between the electron 
current in the TERAS and the average number M, which is shown 
in Fig. 18. Electron currents are shown after electrons were 
injected into the TERAS. While an electron beam current gradually 
decreases due to the collision of electrons with the residual gas 
containing in the ring, the average number $M$ remains almost constant. 

\begin{figure}
\includegraphics[width=.9\linewidth]{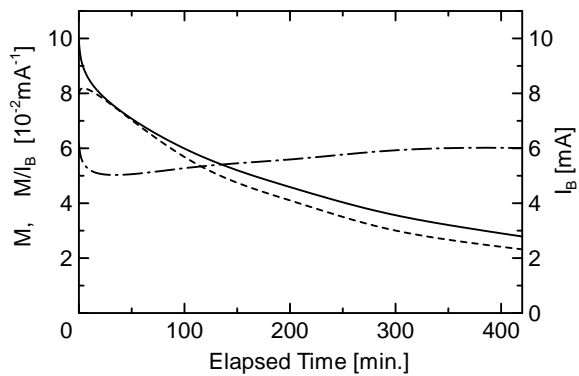}
\caption{\label{fig:Fig18} Time dependences of the electron 
beam current $I_{B}$ (dashed curve), 
the average photon number $M$ (solid curve) and the 
photon production efficiency $M/I_{B}$ (dash-dotted curve) 
for an electron beam current of 1 mA.}
\end{figure}

\subsubsection{\label{sec:level3b4} Energy spectrum of 
incident LCS $\gamma$-ray ($\phi(E_{\gamma})$)}

To determine the photodisintegration cross section of $^{4}$He at a 
certain energy corresponding to an incident LCS $\gamma$-ray, it is 
necessary to measure the intrinsic energy spectrum $\phi(E_{\gamma})$ 
of incident LCS $\gamma$-rays. Note that the LCS $\gamma$-ray has a 
finite energy spread due to the finite widths of the lead collimator 
and of the emittance of electron beams of the TERAS. Hence the LCS 
$\gamma$-ray spectrum was measured using an anti-Compton NaI(Tl) 
spectrometer, which consisted of a central NaI(Tl) detector with a 
diameter of 76.2 mm and a length of 152.4 mm, and an annular one with 
an outer diameter of 254 mm and a length of 280 mm. A typical 
spectrum measured at $E_{max} =$ 28 MeV is shown in Fig. 19. 
Using a response function of the NaI(Tl) detector calculated with 
the GEANT4 simulation code \cite{GEANT4}, the intrinsic energy 
spectrum of the LCS $\gamma$-ray was obtained as shown in Fig. 19. 
An energy spread of LCS $\gamma$-rays was determined as 6\% 
(FWHM) at $E_{max} =$ 28 MeV.

\begin{figure}
\includegraphics[width=.9\linewidth]{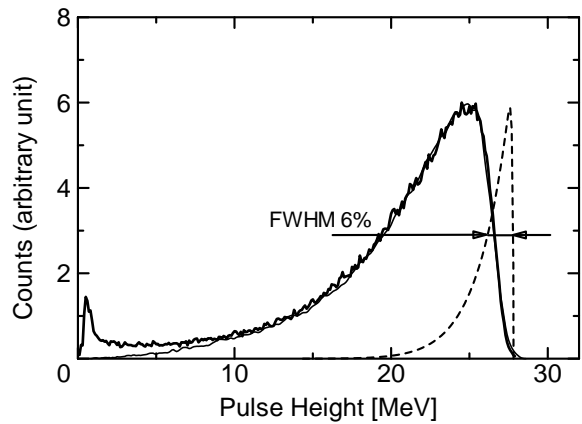}
\caption{\label{fig:Fig19} NaI pulse height spectra obtained 
for $E_{max} =$ 28 MeV. The thick curve and the thin curve 
are the measured one and the calculated one, respectively. 
The dashed curve is an intrinsic energy distribution of the 
LCS $\gamma$-ray required to reproduce the measured 
spectrum.}
\end{figure}

\subsubsection{\label{sec:level3b5}
Photodisintegration of deuteron}

The values of $Y_{i}$, $\varepsilon_{i}$, $N_{t}$, $L$, $\phi$, 
and $\Phi$ were accurately determined as described above, and 
therefore the photodisintegration cross section of $^{4}$He is 
determined accurately using the formula of Eq. 1. It is, however, 
worthwhile to measure the photodisintegration cross section of 
deuteron to learn about any possible systematic uncertainty of 
the present experimental method. Note that the cross section has 
been well studied both experimentally and theoretically in the wide 
energy range from 10 to 75 MeV \cite{Dgpn,Arenhovel}. The present 
measurement was performed using CD$_{4}$ gas instead of CH$_{4}$ 
gas as the quenching gas of the TPC at $E_{max} =$ 22.3 MeV. The 
data were analyzed as extensively described above. The cross 
section turns out to be 0.56$\pm$0.04(stat)$\pm$0.03(syst) mb, 
which agrees nicely with both the previous data \cite{Dgpn} and 
the theoretical value of 0.55 mb \cite{Arenhovel} as shown 
in Fig. 20. The weighted-mean reaction energy was determined 
as being 21.0 MeV using Eq. 3 and the known energy dependence 
of the cross section \cite{Dgpn,Arenhovel}. Hence, the validity 
of the present experimental method including its analysis was 
confirmed with a quite small systematic uncertainty within the 
statistical uncertainty.\\
\newpage

\begin{figure}
\includegraphics[width=.9\linewidth]{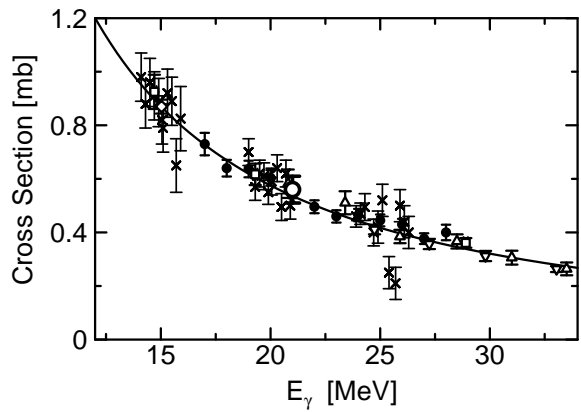}
\caption{\label{fig:Fig20} 
Cross section of the photodisintegration 
of deuteron. The open circle denotes the present result, 
while other symbols indicate the previous data 
\cite{Dgpn}: 
filled circles; Skopik {\it et al.}, 
diagonal crosses; Ahrens {\it et al.}, 
open squares; Bernabei {\it et al.}, 
open diamonds; Michel {\it et al.}, 
open upward triangles; Bosman {\it et al.}, 
open downward triangles; Dupont {\it et al.} 
The solid curve is the theoretical cross section calculated 
by means of the momentum-space approach \cite{Arenhovel}.} 
\end{figure} 

\ \vspace{45mm}\\

\begin{figure}[h]
\includegraphics[width=.9\linewidth]{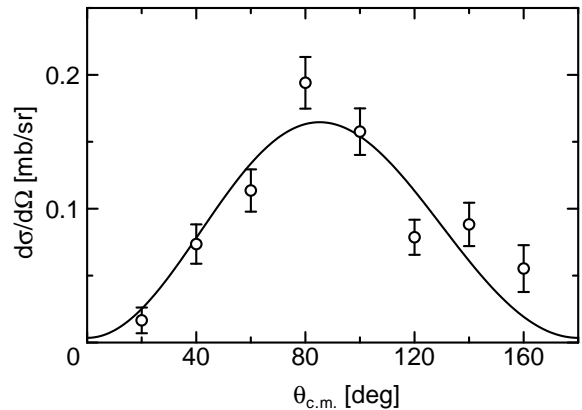} 
\caption{\label{fig:Fig21} Differential cross section of 
the $^{4}$He($\gamma$,p)$^{3}$H reaction at $<E_{\gamma}>=$ 
29.8 MeV. The open circles denote the measured one in the 
present work. The solid curve is the fitted one with the 
least-square method.}
\end{figure}

\ \vspace{10mm}\\

\subsubsection{\label{sec:level3b6}
Angular distribution of proton from the 
$^{4}$He($\gamma$,p)$^{3}$H reaction}

In order to determine the electromagnetic property of the 
photodisintegration at 29.8 MeV we analyzed the angular 
distribution of charged fragments from the 
$^{4}$He($\gamma$,p)$^{3}$H reaction at $E_{max} =$ 32 MeV. 
Note that the two-body total photodisintegration cross 
section below 26.6 MeV is dominated by E1 radiation 
as mentioned above \cite{Ellerkmann}. We analyzed data 
taken at $E_{max} =$ 32 MeV by performing a least-square 
fit to the data using the following formula \cite{Vinokurov}:

\begin{widetext}
\begin{equation}
\frac{d\sigma}{d\Omega} = A_{0} \cdot 
(sin^{2}\theta_{c.m.}+\beta \cdot sin^{2}\theta_{c.m.} \cdot 
cos\theta_{c.m.} +\gamma \cdot sin^{2}\theta_{c.m.} \cdot 
cos^{2}\theta_{c.m.} + \delta + \epsilon \cdot cos\theta_{c.m.})
\label{eq:five}.
\end{equation}
\end{widetext}

Here, $\theta_{c.m.}$ is the angle formed by the proton trajectory 
from the $^{4}$He($\gamma$,p)$^{3}$H reaction with respect to 
the incident LCS $\gamma$-ray beam in the center of mass 
system. $A_{0}$ is determined by the E1 absorption contribution. 
$\beta$ is due to the interference of E1 and E2 electric multipoles, 
$\gamma$ is the ratio of E2 to E1 absorption probabilities, $\delta$ 
is the ratio of M1 to E1 absorption, and $\epsilon$ is an 
isotropic term, which is known experimentally to be approximately 
zero. Consequently, these parameters are 
determined as $A_{0}=$ 0.16$\pm$0.02 mb/sr, $\beta=$ 
0.17$\pm$0.13, $\gamma=$ 0$\pm$0.14, and $\delta=$ 
0.02$\pm$0.01. The results indicate the dominance of an 
electric dipole process in the photodisintegration at 
around 30 MeV, and the M1 strength is about 2\% of the E1 
strength, and the E2 strength is negligible compared to 
the E1 strength. The present result shown in Fig. 21 is in good 
agreement with previous data below 26.2 MeV \cite{Wagenaar} 
and with a theoretical calculation \cite{Ellerkmann}.

\subsubsection{\label{sec:level3b7}
Cross sections of the two-body and three-body 
photodisintegration of $^{4}$He}

In order to calculate average cross sections $<\sigma_{i}>$ of 
the photodisintegration of $^{4}$He and the weighted-mean 
reaction energies $<E_{\gamma}>_{i}$ using Eqs. 2 and 3, we 
first determined the cross sections $<\sigma_{i}(E_{\gamma})>$ 
using the measured yields of the photodisintegration of $^{4}$He 
and Eq. 1 as discussed below. Since we made the measurements at 
four maximum $\gamma$-ray energies $E_{max}$, we can set up 
four simultaneous equations as follows:

\begin{eqnarray}
Y_{i}^{(k)}=N_{t} \cdot L \cdot \Phi \times 
\frac{\int_{0}^{E_{max}^{(k)}} \varepsilon_{i}(E_{\gamma}) 
\cdot \sigma_{i}(E_{\gamma}) \cdot \phi^{(k)}(E_{\gamma}) 
dE_{\gamma}}
{\int_{0}^{E_{max}^{(k)}} \phi^{(k)}(E_{\gamma}) dE_{\gamma}} 
\nonumber \\
(k = 0\sim3) \label{eq:six}.\hspace{6mm}
\end{eqnarray}

Here $k=$ 0, 1, 2 and 3 stand for the measurements at 
$E_{max}=$ 22.3, 25, 28 and 32 MeV, respectively. 
$Y_{i}^{(k)}$ is the measured yield of a reaction channel 
$i$ of the photodisintegration of $^{4}$He in a 
measurement $k$. 
Since  all the resonance states below 30 MeV are 
known to be quite broad \cite{Tilley}, we assume that 
$\sigma_{i}(E_{\gamma})$ is a smooth function of $E_{\gamma}$ 
and practically expressed by a power series of the relative 
momentum $p$ of the particles in the exit channel as 
follows:

\begin{equation}
\sigma_{i}(E_{\gamma}) = \sum_{j=1}^{3} a_{j}p^{j} \ , \ 
p=\left( \frac{\mu (E_{\gamma}-E_{i}^{th})}{2} \right)^{1/2} 
\label{eq:seven}.
\end{equation}

Here $\mu$ and $E_{i}^{th}$ are the reduced mass of the 
emitted particles and the threshold energy in a reaction 
channel $i$, respectively. The coefficients $a_{j}$ were 
determined by solving Eq. 6 and Eq. 7 simultaneously. 
Substituting $\sigma_{i}(E_{\gamma})$ thus obtained for 
Eqs. 2 and 3, the average cross sections $<\sigma>_{i}$ and 
the weighted-mean reaction energies $<E_{\gamma}>_{i}$ were 
obtained. The results for the ($\gamma$,p), ($\gamma$,n), 
and total photodisintegration cross sections are given 
in Table II and shown in Figs. 22(a), 22(b), and 22(c), 
in which the solid curves represent the most 
probable functions obtained with the mentioned procedure 
in the present energy range up to 32 MeV. The systematic 
uncertainties associated with $<\sigma>_{i}$ were 
calculated from the uncertainties in $\varepsilon_{i}$, 
$N_{t}$, $L$ and $\Phi$. Due to the similar excitation 
functions for the $^{4}$He($\gamma$,p)$^{3}$H and 
$^{4}$He($\gamma$,n)$^{3}$He reactions, $<E_{\gamma}>$ 
obtained for both channels agreed with each other within 
100 keV as expected. Hence, we have used the same values of 
$<E_{\gamma}>$ for the ($\gamma$,p) and ($\gamma$,n) 
reaction channels.

\begin{table*}[b]
\caption{\label{tab:table2}
Average photodisintegration cross sections of $^{4}$He. 
The quoted uncertainties are the statistical and systematic 
ones, respectively.}
\begin{ruledtabular}
\begin{tabular}{ccccc}
$<E_{\gamma}>$ [MeV] & $<\sigma(\gamma,p)>$ [mb] 
& $<\sigma(\gamma,n)>$ [mb] & $<\sigma(\gamma,pn)>$ [mb] 
& $<\sigma_{total}>$ [mb] \\ \hline
21.8 & 0.19$\pm$0.02$\pm$0.01 & 0.10$\pm$0.02$\pm$0.006 
& $-$ & 0.29$\pm$0.03$\pm$0.02 \\
24.3 & 0.71$\pm$0.05$\pm$0.03 & 0.63$\pm$0.05$\pm$0.03 
& $-$ & 1.34$\pm$0.07$\pm$0.06 \\
26.5 & 0.89$\pm$0.06$\pm$0.02 & 0.80$\pm$0.06$\pm$0.02 
& $-$ & 1.69$\pm$0.09$\pm$0.04 \\
29.8 & 1.39$\pm$0.08$\pm$0.03 & 1.35$\pm$0.08$\pm$0.03 
& 0.04$\pm$0.01$\pm$0.001 & 2.78$\pm$0.11$\pm$0.06
\end{tabular}
\end{ruledtabular}
\end{table*}

\begin{figure*}[h]
\includegraphics[width=.6\linewidth]{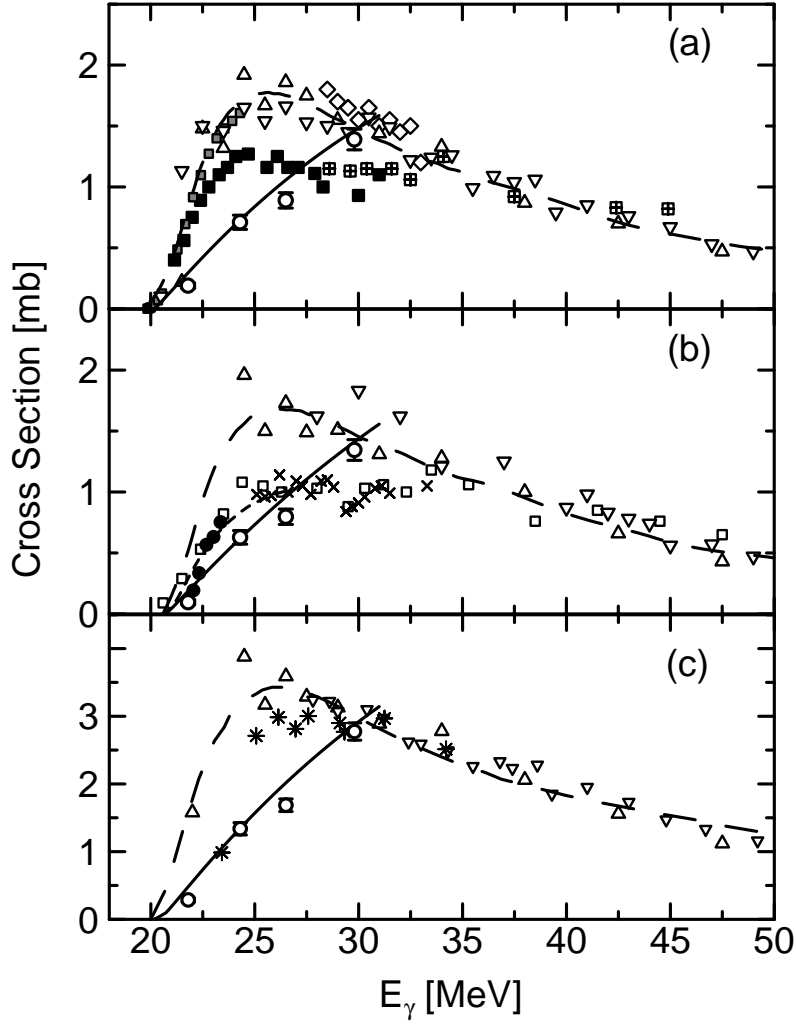} 
\caption{\label{fig:Fig22} $^{4}$He photodisintegration 
cross sections. The solid curves are the most probable 
functions of the cross sections $\sigma_{i}(E_{\gamma})$ 
obtained from the present data in the $\gamma$-ray 
energy range from the reaction threshold up to 32 MeV. 
The open circles denote the average cross sections 
$<\sigma>_{i}$ at the weighted-mean reaction energies 
$<E_{\gamma}>_{i}$, while other symbols indicate 
the previous data: 
(a) ($\gamma$,p) cross sections, 
open upward triangles; Gorbunov \cite{Gorbunov68}, 
open downward triangles; Arkatov {\it et al.} \cite{Arkatov74}, 
crossed squares; Bernabei {\it et al.} \cite{Bernabei88}, 
filled squares; Feldman {\it et al.} \cite{Feldman}, 
open diamonds; Hoorebeke {\it et al.} \cite{Hoorebeke}, 
gray squares; Hahn {\it et al.} \cite{Hahn}, 
(b) ($\gamma$,n) cross sections, 
open upward triangles; Gorbunov \cite{Gorbunov68}, 
open downward triangles; Arkatov {\it et al.} \cite{Arkatov74}, 
open squares; Berman {\it et al.} \cite{Berman}, 
diagonal crosses; Ward {\it et al.} \cite{Ward}, 
filled circles; Komar {\it et al.} \cite{Komar}, 
and (c) total photoabsorption cross sections, 
open upward triangles; Gorbunov {\it et al.} \cite{Gorbunov58}, 
open downward triangles; Arkatov {\it et al.} \cite{Arkatov74}, 
asterisks; Wells {\it et al.} \cite{Wells}. 
The error bars of the previous data are not shown for 
clarity. 
The long-dashed curves are the cross sections calculated 
using the LIT method with the MTI-III potential \cite{Quaglioni}. 
The short-dashed curve represents the calculated ($\gamma$,n) 
cross section based on the AGS formalism \cite{Ellerkmann}.}
\end{figure*}

\section{\label{sec:level4}Discussion}

\subsection{\label{sec:level4a}Ratio of the 
$^{4}$He($\gamma$,p)$^{3}$H cross section 
to the $^{4}$He($\gamma$,n)$^{3}$He cross section}

The cross section ratio 
$R_{\gamma} \equiv \sigma(\gamma,p)/\sigma(\gamma,n)$ has been 
determined accurately with an experimental uncertainty of about 
10\% and with small systematic uncertainties in the energy range 
from 21.8 to 29.8 MeV. The ratio is consistent with calculated 
values without charge symmetry breaking of the strong interaction 
\cite{Unkelbach,Quaglioni,Halderson,Londergan} within the 
experimental uncertainty as shown in Fig. 23. There, the previous 
data taken simultaneously for the 
$^{4}$He($\gamma$,p)$^{3}$H and $^{4}$He($\gamma$,n)$^{3}$He 
reactions are shown for a comparison. Note that the large ratio, 
approximately equal to 2.0 at 21.8 MeV, is due to the difference 
of the Q-values between n-$^{3}$He and p-$^{3}$H channels. The 
ratio at 26.5 and 29.8 MeV agrees with the latest result obtained 
by detecting emitted particles from these reactions simultaneously 
at 90$^{\circ}$ with respect to the incident beam direction. Our 
result is also consistent with simultaneous measurements on the 
$^{4}$He(e,e'X) reaction in the excitation energy region between 
22 and 36 MeV \cite{Spahn} and the $^{4}$He(p,p'X) reaction 
\cite{Raue}. 

\begin{figure}[t]
\includegraphics[width=.9\linewidth]{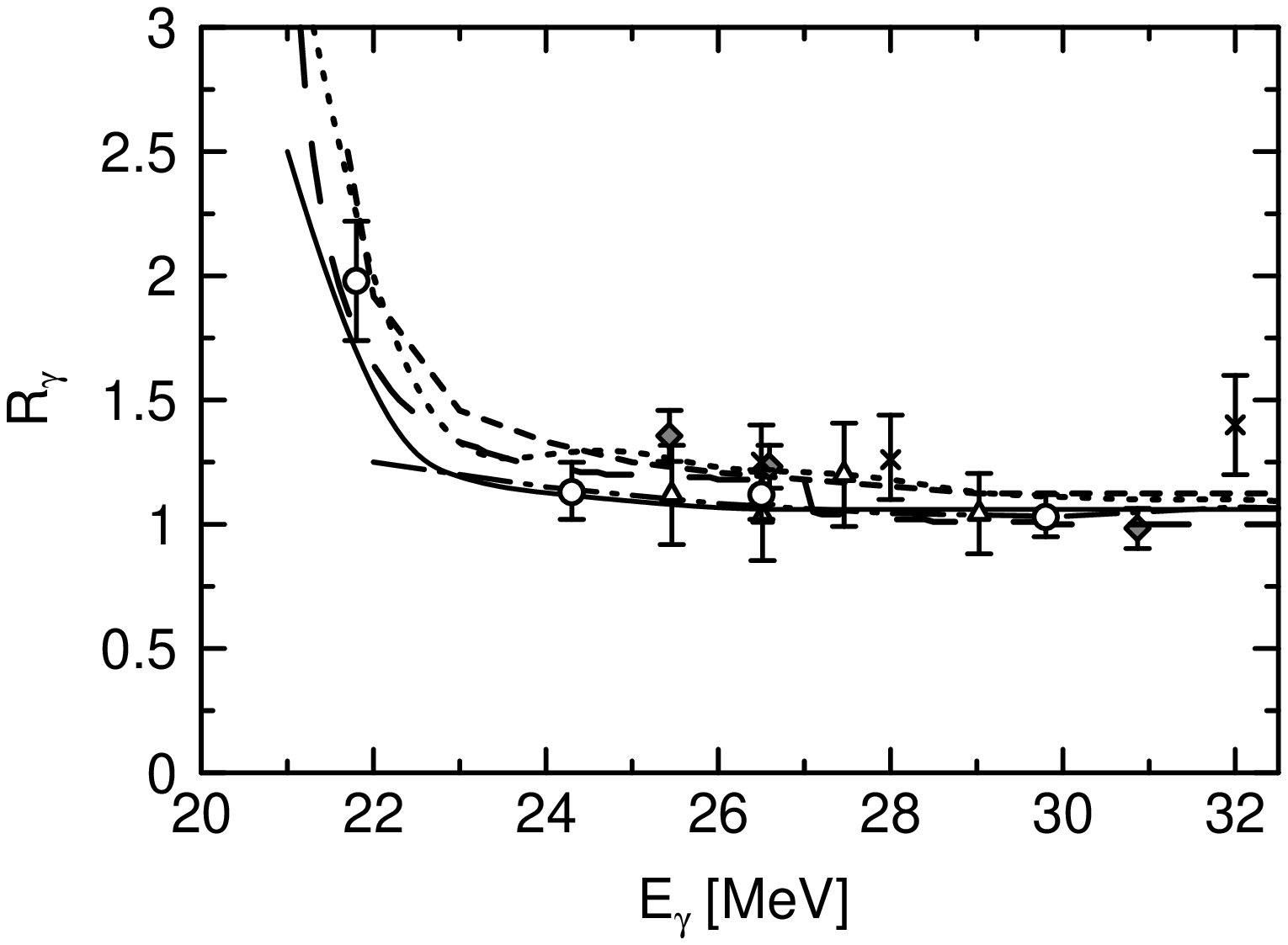} 
\caption{\label{fig:Fig23} Ratio $R_{\gamma}$ of the 
$^{4}$He($\gamma$,p)$^{3}$H cross section to the 
$^{4}$He($\gamma$,n)$^{3}$He cross section: 
open circles; present result, 
open triangles; Gorbunov \cite{Gorbunov68}, 
gray diamonds; Balestra {\it et al.} \cite{Balestra77}, 
diagonal crosses; Florizone {\it et al.} \cite{Florizone}. 
Short-dashed curve and solid curve are the calculations of 
the recoil-corrected continuum shell model with and without 
extra CSB effect, respectively \cite{Halderson}. 
The long-dashed curve, the dotted curve, and the dash-dotted 
curve are the calculations without extra CSB based on 
the LIT method \cite{Quaglioni}, the coupled-channel 
continuum shell model \cite{Londergan}, and the resonating 
group model \cite{Unkelbach}, respectively.}
\end{figure}

\subsection{\label{sec:level4b}The partial and total 
cross-sections of the photodisintegration of $^{4}$He}

The present ($\gamma$,p), ($\gamma$,n), and total 
photodisintegration cross sections of $^{4}$He shown 
in Figs. 22(a), 22(b), and 22(c) differ significantly from 
previous data (see below for details). 

\subsubsection{\label{sec:level4b1}$^{4}$He($\gamma$,p)$^{3}$H}

The $^{4}$He($\gamma$,p)$^{3}$H cross section increases 
monotonically with energy up to 29.8 MeV. The cross section 
below 26.5 MeV does show agreements with none of the previous 
data while at 29.8 MeV it agrees nicely with some of the previous 
data \cite{Gorbunov68,Arkatov74,Hoorebeke} and marginally agrees 
with that by Bernabei {\it et al.} \cite{Bernabei88}. The value 
at 28.6 MeV in Ref. \cite{Bernabei88} is in good agreement with 
an interpolated value of the present data between at 26.5 and 
at 29.8 MeV. Note that the present cross section and excitation 
function significantly differ from the theoretical calculation 
of the LIT method which predicts a pronounced peak at around 
26$-$27 MeV as shown in Fig. 22(a).

\subsubsection{\label{sec:level4b2}$^{4}$He($\gamma$,n)$^{3}$He}

The $^{4}$He($\gamma$,n)$^{3}$He cross section shows similar 
energy dependence as that of the $^{4}$He($\gamma$,p)$^{3}$H 
as shown in Fig. 22(b), and the value up to 26 MeV marginally 
agree with the data of Berman {\it et al.} \cite{Berman} within 
the experimental uncertainty which includes a systematic error 
of 15\%. The cross section at 29.8 MeV is larger than the previous 
data \cite{Berman,Ward} by about 30\% but agrees with the data of 
Gorbunov \cite{Gorbunov68}. The cross section follows the shape 
of the theoretical calculation based on the AGS method up to 
26 MeV, although the experimental value is smaller by about 20\% 
in comparison to the calculation. The experimental result does 
not agree with the calculation based on the LIT method which 
predicts a pronounced peak at around 26$-$27 MeV. 

\begin{figure}[h]
\includegraphics[width=.9\linewidth]{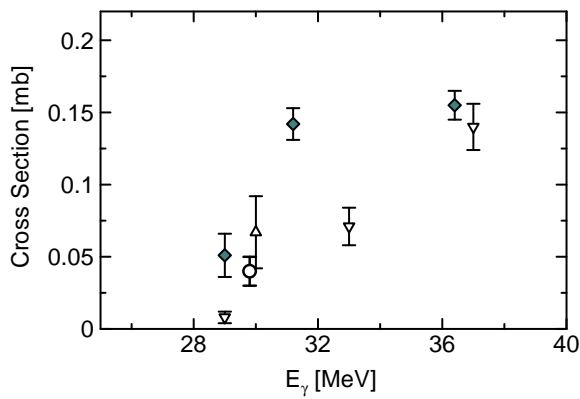} 
\caption{\label{fig:Fig24} $^{4}$He($\gamma$,pn)$^{2}$H 
reaction cross section: open circles; present result, 
open upward triangles; Gorbunov {\it et al.} \cite{Gorbunov58}, 
open downward triangles; Arkatov {\it et al.} \cite{Arkatov70}, 
gray diamonds; Balestra {\it et al.} \cite{Balestra79}.}
\end{figure}

\subsubsection{\label{sec:level4b3}$^{4}$He($\gamma$,pn)$^{2}$H}

The small value of the $^{4}$He($\gamma$,pn)$^{2}$H cross section 
at 29.8 MeV, 0.04$\pm$0.01 mb, agrees with previous data as shown 
in Fig. 24 \cite{Gorbunov58,Arkatov70,Balestra79}. The theoretical 
calculations on the three-body $^{4}$He($\gamma$,pn)$^{2}$H cross 
section are not available. \\
\ \\

\subsubsection{\label{sec:level4b4}Total cross section}

The total cross section increases monotonically with energy up 
to 29.8 MeV as shown in Fig. 22(c). The cross section below 
26.5 MeV is significantly smaller than previous data 
\cite{Gorbunov68,Arkatov74} and a theoretical calculation based 
on the LIT method. The cross section at 29.8 MeV agrees with the 
previous data and with the calculation. Here, it is worthwhile 
to mention that the total photo-absorption cross-section is 
inferred from the elastic photon scattering data of $^{4}$He 
\cite{Wells} together with previous data of the shape of 
the photodisintegration cross section, which claim the GDR peak 
in the region of 25$-$26 MeV. The cross-section inferred turns 
out to be 2.86$\pm$0.12 mb at around 26 MeV \cite{Wells}, which 
differs significantly from the present value of 
1.69$\pm$0.09(stat)$\pm$0.04(syst) mb at 26.5 MeV. The origin 
of the discrepancy is not clear, but it could be due to the 
shape difference between the presently obtained cross section 
and the previous one. Naturally, it would be interesting to 
estimate the total photo-absorption cross-section using the 
shape of the total cross section derived in the present study 
and the photon scattering cross-section data \cite{Wells}. 
Note that the shape can be obtained by combining the present 
results up to 29.8 MeV with the previous data above around 33$-$35 
MeV, where the previous data agree with each other, as shown in 
Fig. 22(c). 

\subsubsection{\label{sec:level4b5}E1 sum rule}

Since the present cross section is found to be smaller than 
many previous data and considering that the E1 transition 
dominates, it is important to investigate the energy distribution 
of the transition strength. It is well known that the integrated 
cross section $\sigma_{0}$ for E1 photoabsorption and the 
inverse-energy-weighted sum $\sigma_{B}$ can be related 
to the properties of the ground state of a nucleus through 
the following sum rules \cite{Levinger,Foldy}:

\begin{widetext}
\begin{eqnarray}
\sigma_{0}= \int_{0}^{E_{\pi}} \sigma_{E1}(E_{\gamma}) dE_{\gamma}
= \sigma_{TRK}(1+\kappa) = \frac{2\pi^{2}e^{2}\hbar}{mc}
\frac{NZ}{A}(1+\kappa) \ , 
\\
\sigma_{B}= \int_{0}^{E_{\pi}} 
\frac{\sigma_{E1}(E_{\gamma})}{E_{\gamma}} dE_{\gamma} 
= \frac{4\pi^{2}}{3}\frac{e^{2}}{\hbar c}\frac{NZ}{A-1} 
\cdot (<r_{\alpha}^{2}> - <r_{p}^{2}>)
\label{eq:8-9}.
\end{eqnarray}
\end{widetext}

Here $\sigma_{E1}(E_{\gamma})$ is the total cross section for 
E1 photoabsorption as a function of $E_{\gamma}$. $N$, $Z$, and 
$A$ are the numbers of the neutrons, protons, and nucleons, 
respectively. $m$ and $\kappa$ are the nucleon mass and the 
correction factor for the contribution of the exchange forces, 
respectively. $E_{\pi}$, $e$, $\hbar$, and $c$ stand for the pion 
threshold energy, electron charge, Planck's constant, and speed 
of light, respectively. $\sigma_{TRK}$ stands for 
the Thomas-Reiche-Kuhn (TRK) sum rule. $<r_{\alpha}^{2}>$ and 
$<r_{p}^{2}>$ are the mean-square charge radii of $^{4}$He and 
the proton, respectively. The integrations in Eqs. 8 and 9 have 
been performed as follows: below 31 MeV we assumed 
$\sigma_{E1}(E_{\gamma})$ is given as the sum of the present 
$\sigma(\gamma,p)$ and $\sigma(\gamma,n)$, because the cross 
section is found to be dominated by the E1 photoabsorption in 
the two-body channels. Above 31 MeV we employed previous data 
of Refs. \cite{Gorbunov68,Gorbunov58} and \cite{Arkatov74,
Arkatov80}, which are in an overall agreement with each other 
as well as with recent theoretical calculations. The $\sigma_{0}$ 
and $\sigma_{B}$ values are listed in Table III. Here it should 
be noted that the data taken from 
Refs. \cite{Gorbunov68,Gorbunov58} and from Refs. 
\cite{Arkatov74,Arkatov80} correspond to the cross sections 
for the total photoabsorption and the E1 photoabsorption, 
respectively. Therefore the data set of the present result and 
the cross section from Refs. \cite{Gorbunov68,Gorbunov58} 
provides upper limits on $\sigma_{0}$ and $\sigma_{B}$. 
The contributions of higher multipoles have been estimated 
to be of a few percent \cite{Gorbunov58,Arkatov80}. 
Consequently, the present value of $\sigma_{0}$ is marginally 
lower than the value expected from the other light nuclei 
\cite{Ahrens} and from theoretically predicted values 
\cite{Efros,Heinze}. As for $\sigma_{B}$, the present value 
is significantly smaller than the calculated value of 
2.62$\pm$0.02 mb obtained from Eq. 8 using the known 
experimental values of 
$<r_{\alpha}^{2}>^{1/2} =$ 1.673$\pm$0.001 fm 
\cite{Borie} and $<r_{p}^{2}>^{1/2} =$ 0.870$\pm$0.008 fm 
\cite{PDG2004}. 

\begin{table*}
\caption{\label{tab:table3} Integrated cross section 
$\sigma_{0}$ and inverse-energy-weighted sum rule 
$\sigma_{B}$ for the E1 photoabsorption of $^{4}$He.}
\begin{ruledtabular}
\begin{tabular}{cccc}
$E_{\gamma}$ [MeV]  & Data set & $\sigma_{0}$ [MeV$\cdot$mb] 
& $\sigma_{B}$ [mb] \\ \hline
19.8$-$31 & Present & 18.1$\pm$2.1 & 0.67$\pm$0.07 \\ \hline
19.8$-$135 & Present $+$ Refs. [25,52]\footnote{Upper limits 
for E1 contribution.} 
& 96$\pm$7 & 2.24$\pm$0.17\\ \cline{2-4}
\  & Present $+$ Refs. [26,53] & 80.4$\pm$2.3 
& 1.92$\pm$0.12 \\ \hline
Sum rule (see text.) & \ & 100$\sim$128 & 2.62$\pm$0.02 
\end{tabular}
\end{ruledtabular}
\end{table*}

\subsubsection{\label{sec:level4b6}Present results and previous data}

In the present simultaneous measurement of the 
$^{4}$He($\gamma$,p)$^{3}$H and $^{4}$He($\gamma$,n)$^{3}$He 
cross-sections using the 4$\pi$ TPC we could get the cross-section 
ratio $R_{\gamma}= \sigma(\gamma,p)/\sigma(\gamma,n)$ with smaller 
systematic uncertainties, which is consistent with the results 
obtained by other reactions as well as with recent simultaneous 
measurement \cite{Florizone}. However, there are discrepancies between 
the present two-body and total cross sections and previous ones as 
described before. Although it is difficult to find a unique reason of 
the discrepancies since a special care has been taken for the 
normalization of each of the cross section measured here, it might be 
instructive to look at general trends recognized in the previous data 
and to compare them to the present results. 

Firstly we discuss the latest $^{4}$He($\gamma$,p)$^{3}$H data using 
bremsstrahlung photons \cite{Hoorebeke}, which are larger than the data 
\cite{Bernabei88} by about 40\% at around 30 MeV and are in good 
agreement with the data \cite {Bernabei88} at 33 MeV within the 
experimental uncertainty. Note that the data \cite{Bernabei88} were 
obtained using a monochromatic photon beam. This fact may indicate that 
a possible origin of the above mentioned discrepancy is due to 
background events inherent to measurement with bremsstrahlung photons. 
The former group used Si(Li) detectors to detect protons from the 
$^{4}$He($\gamma$,p)$^{3}$H reaction. They carefully considered possible 
systematic errors such as the $^{4}$He gas purity, the gas pressure, 
the efficiency of the Si(Li) detectors, the energy losses in the gas 
target, the incident flux calibration, the bremsstrahlung shape, the 
effects due to the background subtraction and others. They had to 
subtract the background component in the Si(Li) detector assuming an 
exponential fit to the low-energy photon data. They claimed that the 
validity of their background subtraction method has been supported by 
the test measurement of the $^{16}$O($\gamma$,p)$^{15}$N experiment. 
However, the cross section of the $^{16}$O($\gamma$,p)$^{15}$N reaction 
is about 10 times larger than that of $^{4}$He, and therefore the 
ambiguity due to the background subtraction might not have been relevant 
in the test experiment. The contribution of background due to 
bremsstrahlung photons is expected to decrease with increasing the 
photon energy, since the energy of an emitted proton becomes higher. 
Consequently the energy dependence of the difference between the two 
data sets mentioned above can be explained in this way. 

Secondly, we discuss the result of the $^{4}$He($\gamma$,n)$^{3}$He 
reaction obtained by Berman {\it et al.} \cite{Berman}, which was 
carried out using annihilation photon beams and BF$_{3}$ tubes embedded 
in a paraffin matrix as a neutron counter. They carefully made various 
corrections due to the background from bremsstrahlung photons, the neutron 
detector efficiency, and others, and they concluded that their data points 
at 25.3, 26.3 and 28.3 MeV should have systematic uncertainties as 
large as 15\%. If we take the systematic error in addition to statistical 
error, our data marginally agree with the data by Berman {\it et al.} 
It should be stressed that we used a quasi-monoenergetic pulsed Laser 
Compton backscattering (LCS) photon beam, which is free from background 
inherent to bremsstrahlung photon beams, and we detected $^{3}$He 
unambiguously by the nearly 4$\pi$ TPC containing $^{4}$He gas as an 
active target. Hence, we could determine the detection efficiency of 
$^{3}$He with high accuracy. 

\section{\label{sec:level5}Conclusion}

In the present work, we have carried out for the first time 
the direct simultaneous measurement of the two-body and 
three-body photodisintegration cross sections of $^{4}$He in the 
energy range from 21.8 to 29.8 MeV using a quasi-monoenergetic pulsed 
real photon beam by detecting a charged fragment with a nearly 4$\pi$ 
time projection chamber in an event-by-event mode. The validity of 
the present new experimental method, including its data analysis, 
has been accurately confirmed by measuring the photodisintegration 
cross section of deuteron. By accurately determining the ratio 
of the $^{4}$He($\gamma$,p)$^{3}$H to $^{4}$He($\gamma$,n)$^{3}$He 
cross sections, we have solved for the first time the long-standing 
problem of the large discrepancy in this ratio obtained in separate 
measurements and simultaneous ones. The $^{4}$He($\gamma$,p)$^{3}$H, 
$^{4}$He($\gamma$,n)$^{3}$He and total cross sections do not agree 
with the recent calculations based on the Lorentz integral transform 
method. The $^{4}$He($\gamma$,n)$^{3}$He cross section follows the 
shape of the calculation based on the AGS method up to 26.5 MeV, 
but it is smaller by about 20\% with respect to the calculated values. 
We conclude that further theoretical work in the GDR energy region 
is necessary to elucidate the GDR property of $^{4}$He. 
Concerning the photonuclear reactions of three-nucleon systems, it 
has been known that 3NF reduces the peak cross section by about 
10$-$20\% \cite{Skibinski}. Since $^{4}$He is tightly bounded 
compared to the three-nucleon systems, one might expect significant 
3NF effects in the photodisintegration of $^{4}$He. The present result 
would affect significantly the production yields of r-nuclei by 
the neutrino-induced r-process nucleosynthesis, since the 
neutral current neutrino spallation cross sections are quite 
sensitive to the peak energy of the GDR and the cross section 
in the GDR energy region. 

\begin{acknowledgments}
We would like to thank Profs. H. Kamada and T. Kajino for 
discussions, and Dr. A. Mengoni for comments and careful reading of 
the manuscript. The present work was supported in part by 
Grant-in-Aid for Specially Promoted Research of the Japan 
Ministry of Education, Science, Sports and Culture and in 
part by Grant-in-Aid for Scientific Research of the Japan 
Society for the Promotion of Science (JSPS). 
\end{acknowledgments}

\end{document}